\documentclass[aps,prb,reprint,groupedaddress]{revtex4-2}

\usepackage{graphicx}
\usepackage{color,soul}
\usepackage{amssymb}
\usepackage[caption=false]{subfig}
\usepackage{xcolor}
\usepackage{amsmath,upgreek}
\usepackage{wasysym}

\setcitestyle{super}

\begin{document}

\title{Defect-mediated orientational ordering in stripe-forming magnetic systems}

\author{Premarun Barik}
\email{barikp@myumanitoba.ca}
\affiliation{Department of Physics and Astronomy, University of Manitoba, Winnipeg, MB, R3T 2N2, Canada}

\author{Johan van Lierop}
\affiliation{Department of Physics and Astronomy, University of Manitoba, Winnipeg, MB, R3T 2N2, Canada}
\email{Johan.van.Lierop@umanitoba.ca}

\author{Robert L. Stamps}
\affiliation{Department of Physics and Astronomy, University of Manitoba, Winnipeg, MB, R3T 2N2, Canada}
\email{Robert.Stamps@umanitoba.ca}

\date{\today}

\begin{abstract}

We investigate the nonequilibrium development of orientational order in a two-dimensional stripe-forming magnetic system with competing interactions. Following a quench into the stripe-ordered regime, horizontal and vertical stripes organize into orientational regions(super-domains) separated by characteristic defects. We show that the evolution towards a globally oriented stripe state is mediated by the motion and elimination of these defects, with T-junctions providing a direct microscopic measure of super-domain reorganization. The T-junction population reveals distinct dynamical regimes associated with super-domain formation, growth, and equilibration. Temperature, system size, and equilibrium stripe width modify the characteristic time scales of this evolution, while thermal fluctuations produce a finite defect population at higher temperatures. We further show that the defect dynamics observed in real space are reflected in the structure factor, providing a reciprocal-space signature of defect elimination. These results establish a microscopic defect-based description of orientational ordering in stripe-forming magnetic systems.

\end{abstract}


\maketitle


\section{Introduction\label{sec:intro}}

Pattern formation is common in systems when interactions between atoms or molecules compete over different length scales. While short-range interactions favour local order, long-range interactions can oppose this order, and such systems typically form modulated patterns or structures, as they are the lowest energy configurations. Depending on the relative strengths of competing interactions, and on external influences (e.g. electric and magnetic fields), the penultimate patterns have been found to be bubbles, labyrinthine stripe patterns, or periodic stripe patterns\cite{kooy1960experimental, 10.1063_1.328121, 8a265d0d-ebde-3389-9ee4-4c2c542a04ef, PhysRevE.111.055406, tripathi2019coarsening}.  Periodic stripe patterns encapsulate all the underlying physics without an external field applied and have the simplest resultant patterns. Stripe patterns are most easily created in thin films with a perpendicular magnetic anisotropy.  Examples are ultrathin magnetic films and multilayers of LiHoF$_4$\cite{10.1117_12.961386, battison1975ferromagnetism}, Co/Pd\cite{10.1063_1.357538} and Co/Pt\cite{LOUAIL1997387}. Stripe-domain structures are also found in other ferroic materials like ferroelectric PbTiO$_3$ thin films\cite{tovaglieri2026superdomains} where the stripes can exhibit preferred orientations along crystallographic directions\cite{simmons2024ferromagnetic, 10.1117_12.961386}.  Overall, it is found that the short-range exchange interactions favour parallel alignment of neighbouring moments while long-range dipolar interactions compete with the formation of a uniform magnetization or polarization.  The balance between these interactions produces alternating regions of magnetization (polarization) pointing perpendicular to the film plane, resulting in stripe-domain patterns. 

These experimental systems illustrate that, in addition to the formation of a stripe pattern, the orientation of the stripes constitutes an important aspect of the lowest energy configuration. However, the evolution of these stripes towards their final configuration is still an open question. More generally, the evolution of domains following a quench remains an active problem across a wide range of ordered systems, including liquid crystals, active matter, block copolymers, and other symmetry-broken systems. Recent studies have shown that the resulting ordering kinetics can depend strongly on the motion, interaction, and elimination of domain boundaries and localized defects\cite{huh2024universality, fumeron2023introduction, angheluta2026full, balch2023spatially, singh2021late, liang2026phase, indergand2023domain, pinna2025mechanisms, patel2022rapid, leniart2022pathway, blagojevic2023multiscale, mondal2024ordering}. Although the specific defects differ from one system to another, the same basic question remains: which microscopic structures move the boundaries and allow one type of domain to grow at the expense of another?

We do that using a model that captures the essential physics of stripe formation using a perpendicular-anisotropy magnetic film system described by a two-dimensional Ising model.  This model has short-range exchange and long-range dipolar interactions\cite{PhysRevB.51.16033, arlett1996phase}, providing a clear representation of the competing interactions. Prior research has found that the local exchange favours the alignment of neighbouring spins, and the dipolar interactions frustrate this alignment to create modulated structures. The model reproduces several important features of stripe-domain systems and has therefore been used extensively\cite{de2020emergent}. Previous simulations have focused on the equilibrium properties such as how the final stripe width depends on the relative strengths of exchange and dipolar interactions\cite{booth1995domain, arlett1996phase}, how the stripe structure evolves with temperature\cite{whitehead2008canted}, how an applied field modifies the equilibrium patterns\cite{arlett1996phase}, and how a system undergoes transitions between stripe-ordered and disordered phases\cite{gabay1985phase, rizzi2010phase, principi2016stripe}. The local spin behaviour that leads to stripe formation near the transition temperature has also been investigated using quantities such as the orientational order parameter\cite{bab2019evidence, horowitz2015phase}.

To date, the nature of the stripe evolution is understood to change depending on how the system is quenched\cite{tripathi2015coarsening, gleiser2003slow}. If the quench is over a large enough temperature range, there have been hints that the stripe evolution is governed less by critical fluctuations and more by the overall reorganization of the emerging stripe pattern. This raises an important question: what microscopic processes are responsible for the development of a single global stripe orientation?

In part, this question arises because a global order parameter such as the orientational order parameter does not capture how intermediate stripe domains reorganize into a single orientational state. For example, on a square lattice, horizontal and vertical stripe orientations are equivalent. After a quench from a disordered state, locally ordered stripe domains form in both directions. The presence of these differently oriented domains does not by itself determine the eventual global orientational order. The system must reorganize these regions until super-domains\cite{gleiser2003slow, sampaio1996magnetic, hosiawa2009late, cannas2008interplay} form and subsequently evolve toward a single global stripe orientation. This leads to a second question: how does this reorganization occur?

To address these questions, it is necessary to examine the evolution of real-space spin configurations.  In agreement with previous works, we find that stripes evolve into horizontal and vertical stripe domains\cite{gleiser2003slow, sampaio1996magnetic, cannas2008interplay, hosiawa2009late, bromley2003memory}.  As in experiments, we observe T-junctions\cite{sun2024reversible, saxena2025strain, moon2019measuring, singh2018nucleation, shi2023domain}, but also identify corners and I-junctions at the boundaries between differently oriented stripe domains. These junctions are defined later in Fig.~\ref{fig:VerticalHorizontalSuperdomains}. These corners and junctions are defects whose motion is tied to the growth and shrinkage of differently oriented stripe domains via, e.g. junction interactions and their elimination. In particular, the dynamics of the T-junctions provide a direct microscopic connection between the evolution of domains, super-domains and the emergence of a global orientational order (fully aligned spin configuration) without an external driving field.  We are able to describe the ordering processes in terms of the defect populations and their evolution (in Monte Carlo time). Tracking the defects enables a microscopic picture of how a system with two equivalent stripe domains selects a single orientation.


\section{Model and simulations}

We use a two-dimensional Ising model with competing exchange and dipolar interactions on a square lattice. The Hamiltonian is\cite{PhysRevB.51.16033}
\begin{equation}
H = -\delta \sum_{\langle i,j\rangle} S_i S_j
+
\sum_{i<j} \frac{S_i S_j}{r_{ij}^{3}},
\end{equation}
where $S_i=\pm 1$ represents the Ising spin at lattice site $i$. The first term describes a ferromagnetic nearest-neighbour exchange interaction. The sum, $\langle i,j\rangle$, runs over nearest-neighbour pairs. The second term in $H$ represents the long-range dipolar interaction between all pairs of spins. Here, $r_{ij}$ is the distance between spins located at sites $i$ and $j$, measured in units of the lattice spacing. $\delta$ reflects the relative strength of the exchange relative to dipolar interactions were positive $\delta$ favours parallel alignment of neighbouring spins.  The competition between these two interactions produces stripe configurations. The parameter $\delta$ controls the width of the stripe patterns. Increasing $\delta$ strengthens the alignment between neighbouring spins, which increases the equilibrium stripe width.


\begin{figure}[t!]

\centering

\includegraphics[width=0.475\textwidth]{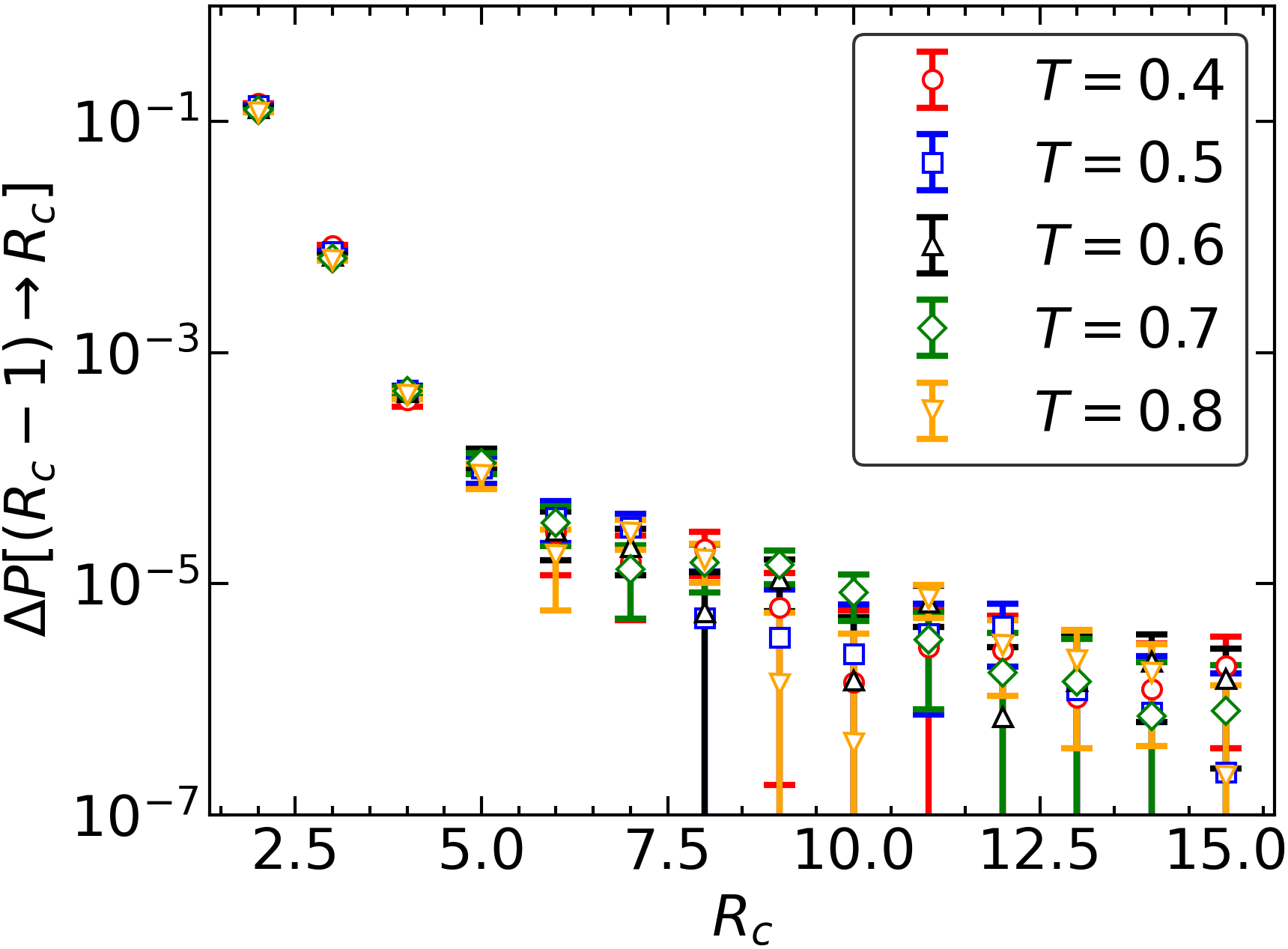}

\caption{Convergence with dipolar-interaction cutoff. Change in the single-spin transition probability, $\Delta P = P(R_c) - P(R_c + 1)$, when the dipolar cutoff is increased by one lattice spacing. Results are averaged over 1000 independent configurations and are shown for several temperatures. The magnitude of $\Delta P$ decreases rapidly with $R_c$ and is of order $10^{-5}$ between $R_c = 10$ and $11$, indicating that contributions from spins beyond this distance produce only a small correction to the transition probability.\label{fig:DipolarCutoffTransitionProbability}}

\end{figure}


For the dipolar interactions, evaluating the interactions between every pair of spins becomes computationally expensive as the system size increases\cite{ewald1921ewald, lekner1989summation}. To reduce the computational cost, we introduce a finite cut-off radius, $R_c$, beyond which the dipolar interaction was neglected. The value of $R_c$ was chosen by examining how the transition probability for a spin-flip changed as the cut-off radius increased.  This transition probability is defined as $P = \exp{(-\Delta(E)/k_{\text{B}}T)}$, where $\Delta(E)$ is the change in the total interaction energy from dipolar and exchange energies associated with a spin flip at temperature $T$, and $k_{\text{B}}$ is Boltzmann's constant. $\Delta P$ is the change in the transition probability when the dipolar cut-off radius is increased by one lattice spacing.  Figure~\ref{fig:DipolarCutoffTransitionProbability} shows $\Delta P$ as a function of $R_c$, averaged over 1000 independent replicas. As the cut-off radius increases, $\Delta P$ decreases rapidly and becomes $\sim 10^{-5}$ when $R_c=10$, which we used in the simulations as $R_c > 10$ produced no measurable effects on $\Delta P$.


\begin{figure*}[t]

\centering

\includegraphics[width=0.875\textwidth]{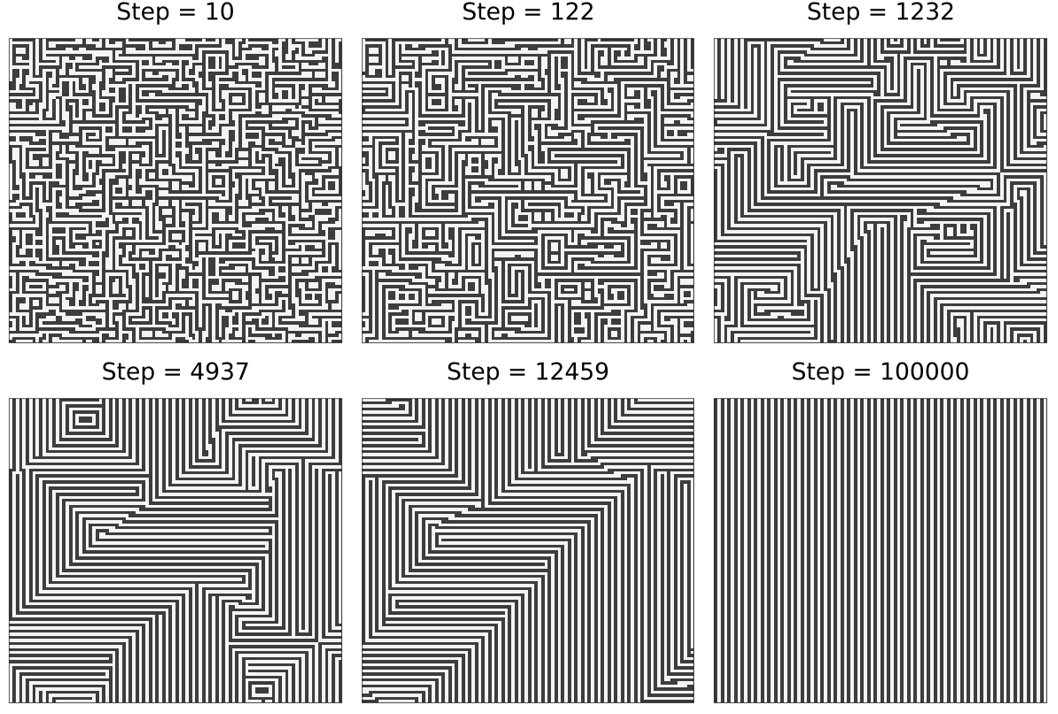}

\caption{Representative snapshots of stripe evolution following a quench from a random configuration to $T=0.5$. Local horizontal and vertical stripe regions grow with time, eventually producing a globally ordered stripe state. Snapshots are shown at $t$=10, 122, 1232, 4937, 12459, and  100000 Monte Carlo steps.\label{fig:StripeEvolutionPathway}}

\end{figure*}


Periodic boundary conditions were applied for the nearest-neighbour exchange interaction. This removed edge effects associated with the finite simulation lattice size, the usual approximation of a real system.  The system was initialized in a completely random spin configuration. It was then quenched instantaneously to a temperature below the stripe-ordering temperature, and the evolution of stripes was examined using a Markov-chain Monte Carlo procedure based on single-spin-flip updates. At each update attempt, a spin is selected and flipped according to the corresponding transition probability determined by the change in energy, where a Monte Carlo step is defined as when the attempted spin flips equal the total number of spins in the system.  Thus, on average each spin is selected once per Monte Carlo step.

We ran simulations with exchange-to-dipolar interaction ratios of $\delta=1.1$, $2.0$, and $2.5$ corresponding to equilibrium stripe widths of 1, 2, and 3 lattice spacings, respectively. In agreement with the literature, we found the corresponding stripe-ordering transition temperatures $T_c \simeq 0.38, 0.8$, and $1.06$\cite{bab2019evidence, horowitz2015phase}.  For each $\delta$, the quench temperature was below $T_c$ to ensure the system evolved within the stripe-ordered regime and to allow tracking of defect dynamics with quench temperature and stripe width.


\section{Results and discussion}

The evolution toward global stripe order begins with the formation and reorganization of locally oriented stripe regions. To understand how this process occurs, we first examine the real-space configurations that develop after the quench and identify the characteristic structures that appear at the boundaries between differently oriented regions. This provides the basis for describing the subsequent ordering process in terms of defect motion and elimination.


\subsection{Super-domains and junction defects}

The evolution of the stripe pattern following a quench is shown in Fig.~\ref{fig:StripeEvolutionPathway}. At early times, the system contains short and irregular stripe segments with no preferred global orientation. As the system evolves in time (measured by the number of Monte Carlo steps), these local stripe segments organize into larger regions in which the stripes are aligned either horizontally or vertically, known as super-domains\cite{gleiser2003slow, sampaio1996magnetic, hosiawa2009late, cannas2008interplay}.

\begin{figure}[b!]

\centering

\subfloat[\label{fig:HorizontalSuperDomain}]{\includegraphics[width=0.225\textwidth]{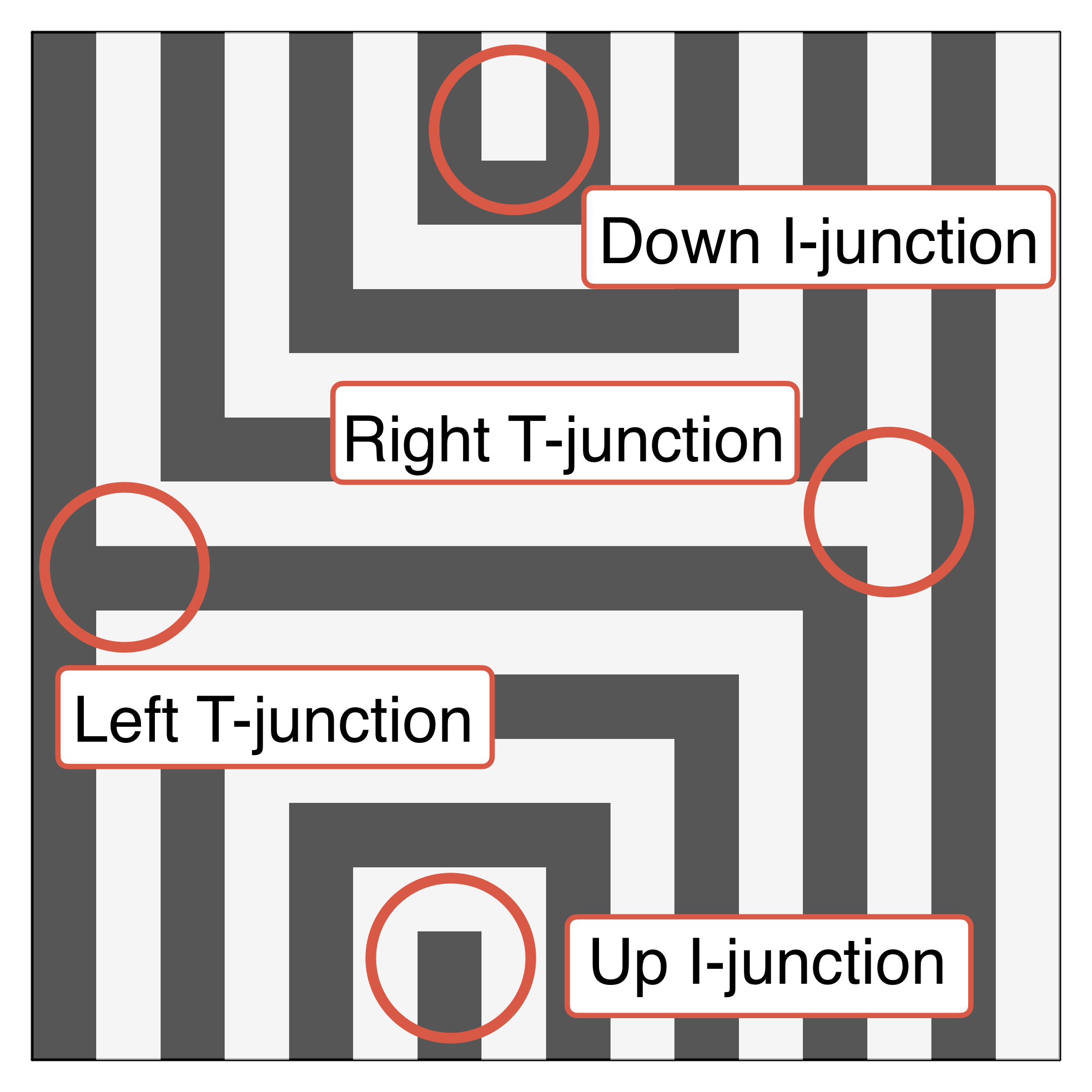}}
\hfill
\subfloat[\label{fig:VerticalSuperDomain}]{\includegraphics[width=0.225\textwidth]{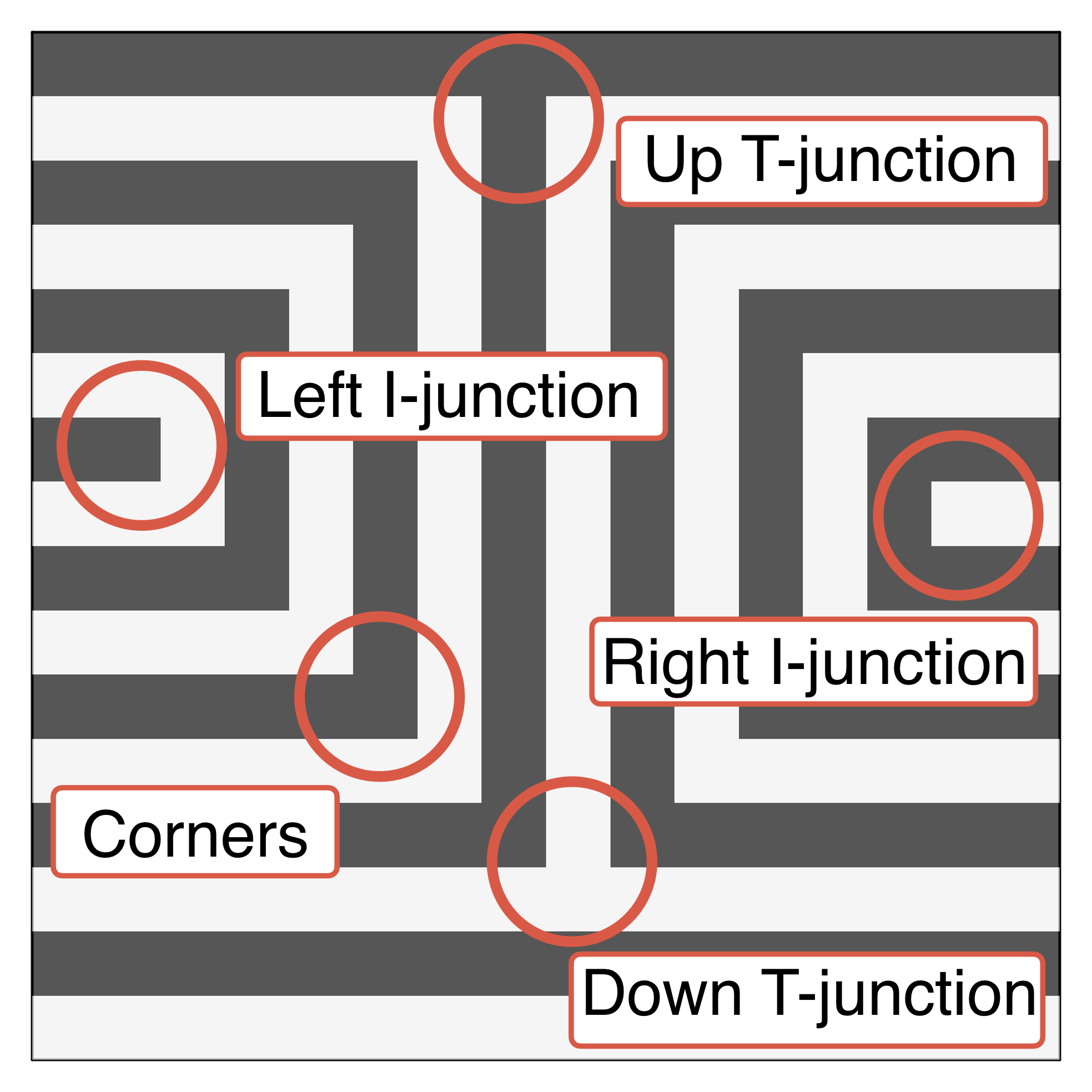}}

\caption{Schematic illustration of the different defect types associated with orientational super-domains. (a) Horizontal super-domain embedded in a vertical-stripe background. The boundary contains left and right T-junctions together with up and down I-junctions. (b) Vertical super-domain embedded in a horizontal-stripe background. In this case, the boundary contains up and down T-junctions together with left and right I-junctions. Representative corners along the super-domain boundary are also indicated.\label{fig:VerticalHorizontalSuperdomains}}

\end{figure}

Because the horizontal and vertical stripe orientations are energetically equivalent, both types of super-domains can form following the quench. The system must therefore reorganize these differently oriented regions before a single stripe orientation can dominate globally.  The boundaries separating super-domains of different orientations follow the diagonal elements of the square lattice, forming at $45^\circ$ and $135^\circ$\cite{bromley2003memory}. Along these boundaries, the local stripe structure gives rise to characteristic defects classified as corners, T-junctions, and I-junctions. Representative examples of these structures are shown schematically in Fig.~\ref{fig:VerticalHorizontalSuperdomains}, where corners correspond to local changes in the direction of the super-domain boundary, and T-junctions and I-junctions occur where the stripe patterns terminate or connect at the interface between the two spin orientations.


The T- and I-junctions are related directly to the orientation of the associated super-domain, illustrated in Fig. \ref{fig:HorizontalSuperDomain}. A horizontal super-domain embedded in a vertical-stripe background contains horizontal T-junctions (left-right) and vertical I-junctions (up-down). Conversely, a vertical super-domain embedded in a horizontal-stripe background contains vertical T-junctions (up-down) and horizontal I-junctions (left-right), as shown in Fig.~\ref{fig:VerticalSuperDomain}. The defect orientations provide a clear marker of the orientation of the corresponding super-domain.  I-junction pairs identify the same trends and behaviours as for a given super-domain, where T-junctions occur on one pair of opposite sides, and I-junctions occur on the other pair.


\begin{figure}[t!]

\centering

\includegraphics[width=0.475\textwidth]{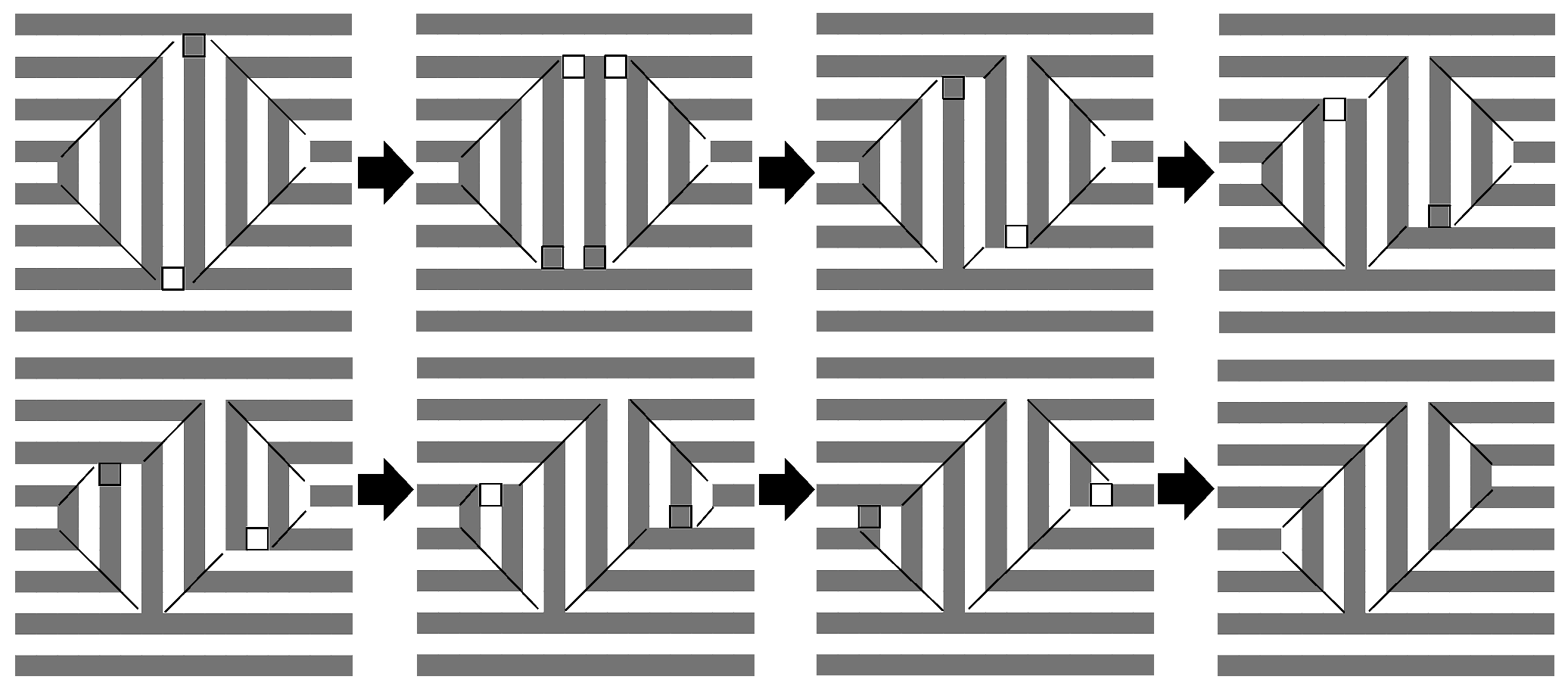}

\caption{A schematic illustration of T-junction motion leading to the shrinkage of a vertical diamond-shaped super-domain. The small boxes indicate regions where spin flips are most probable, resulting in successive displacement of the T-junctions and reduction of the super-domain area. Black lines are included as guides to the eye to identify the super-domain boundary.\label{fig:T-junctionMovementPathway}}

\end{figure}


Among these defects, T-junctions are particularly useful for describing the subsequent evolution because their displacement is associated directly with motion of the super-domain boundaries. As a T-junction moves through successive local spin rearrangements, boundaries advance or retreat with changing size and shape of the corresponding super-domain. This process is illustrated schematically in Fig.~\ref{fig:T-junctionMovementPathway} for a vertical super-domain embedded in a horizontal-stripe background. The vertical T-junctions located along the boundary move toward one another as favourable spin flips occur near the junctions. Their motion corresponds to the area occupied by the vertical super-domain shrinking while the surrounding horizontal stripe region expands. Repeated displacements of T-junctions therefore lead to the progressive shrinking of the vertical super-domain and, eventually, its disappearance.  An equivalent process occurs for a horizontal super-domain embedded in a vertical-stripe background, now described by the horizontal T-junctions.

\begin{figure}[b!]

\centering

\subfloat[\label{fig:TJunctionCountVerticalDominatingReplicas}]{\includegraphics[width=0.475\textwidth]{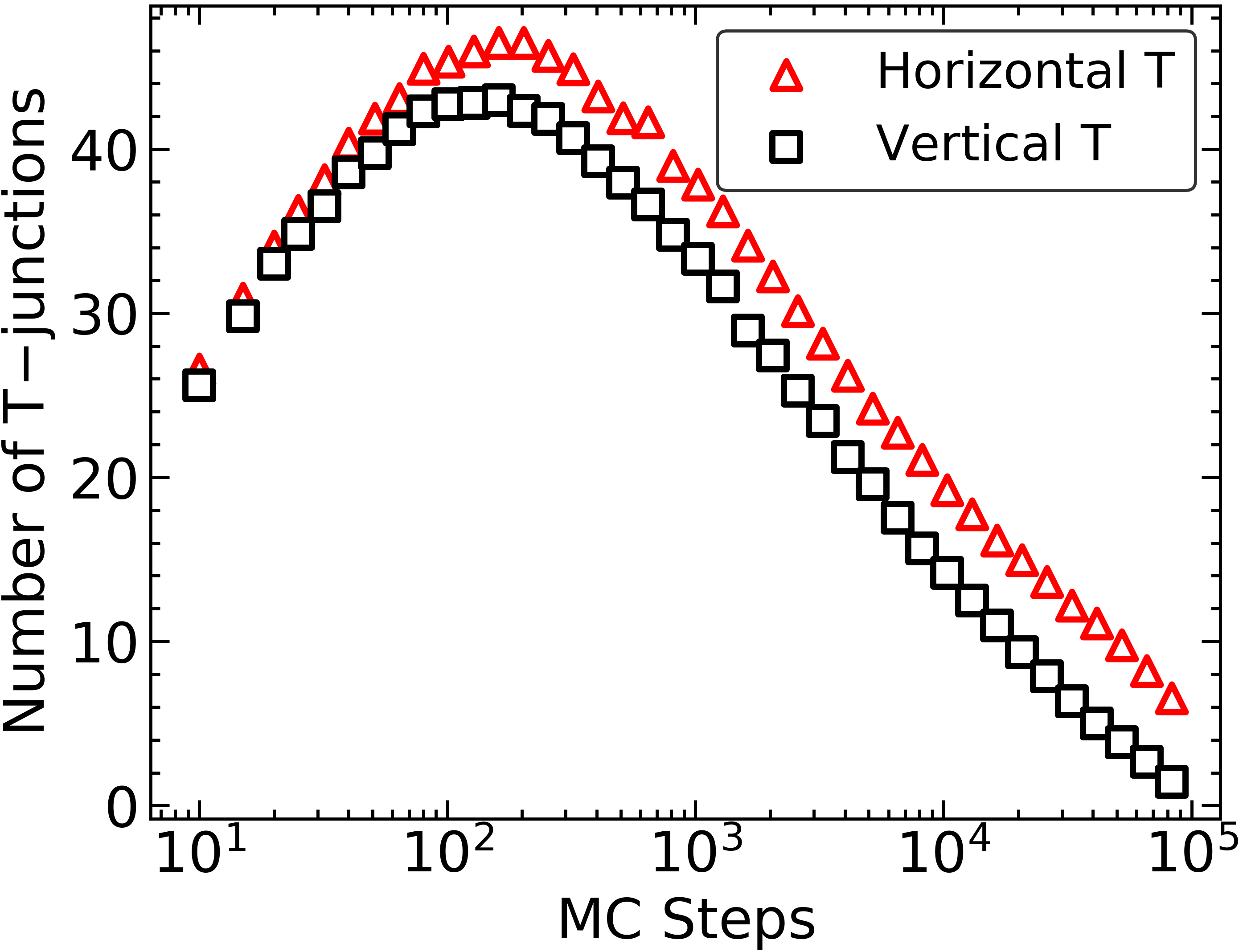}}

\subfloat[\label{fig:TJunctionCountHorizontalDominatingReplicas}]{\includegraphics[width=0.475\textwidth]{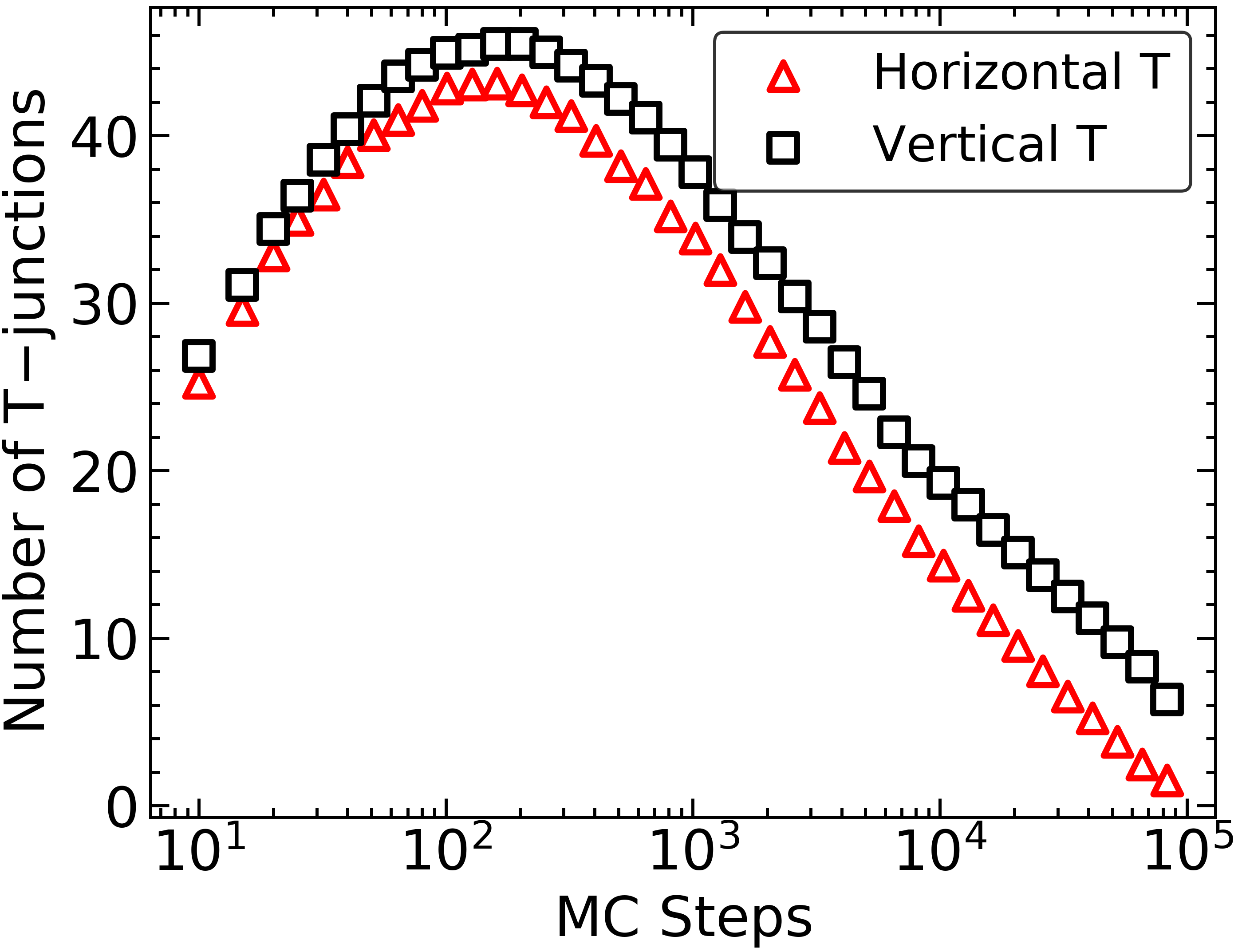}}

\caption{T-junction populations collected from an evolving stripe system. Average numbers of horizontal and vertical T-junctions as functions of Monte Carlo time for 3000 independent $128 \times 128$ systems with $\delta=2.0$ quenched to $T=0.5$. (a) Replicas in which vertical stripes become dominant show a larger population of horizontal T-junctions. (b) Replicas in which horizontal stripes become dominant show the opposite imbalance. The reversal of the defect populations is consistent with horizontal T-junctions mediating the evolution of horizontal super-domains and vertical T-junctions mediating the evolution of vertical super-domains.\label{fig:HorizontalVerticalTJunctionCount}}

\end{figure}

The imbalance between vertical and horizontal T-junctions should be related to the stripe orientation that eventually becomes dominant. Vertical T-junctions are associated with shrinking vertical super-domains and horizontal T-junctions with shrinking horizontal super-domains. Therefore, the dominant stripe orientation should be accompanied by an excess of T-junctions associated with the competing orientation.  These trends are presented in Fig.~\ref{fig:HorizontalVerticalTJunctionCount}, which describes the evolution of vertical and horizontal T-junctions averaged over 3000 independent $128 \times 128$ replicas for $\delta=2.0$ quenched to $T=0.5$. 

The replicas were separated according to the stripe orientation that dominates at late times.  For replicas that ultimately develop a vertically ordered stripe state, the number of horizontal T-junctions remains larger than the number of vertical T-junctions, as shown in Fig.~\ref{fig:TJunctionCountVerticalDominatingReplicas}; the remaining horizontal super-domains are being eliminated, and the defects associated with their boundaries are dominant. Conversely, for replicas that evolve toward a horizontally ordered stripe state, the vertical T-junction population becomes larger (Fig.~\ref{fig:TJunctionCountHorizontalDominatingReplicas}) because vertical super-domains constitute the competing regions that must shrink and disappear.  Clearly, T-junction populations provide a microscopic indicator of the path to global orientational order.


\subsection{Defect-mediated orientational ordering}
Having made a defect-based description of the full ordering process, we turn to the simplest stripe configuration: a stripe width of one $(h = 1)$. The evolution of stripes is characterized via complementary quantities of i) total T-junction populations that track the overall number of orientational defects, and ii) the spatial separation between T-junctions that provides information about the characteristic size of the associated stripe regions. Both i) and ii) quantify how the overall system approaches equilibrium through super-domains.


\begin{figure}[t!]

\centering

\subfloat[\label{fig:T-junctionCountTimeDependanceT0.1}]{\includegraphics[width=0.475\textwidth]{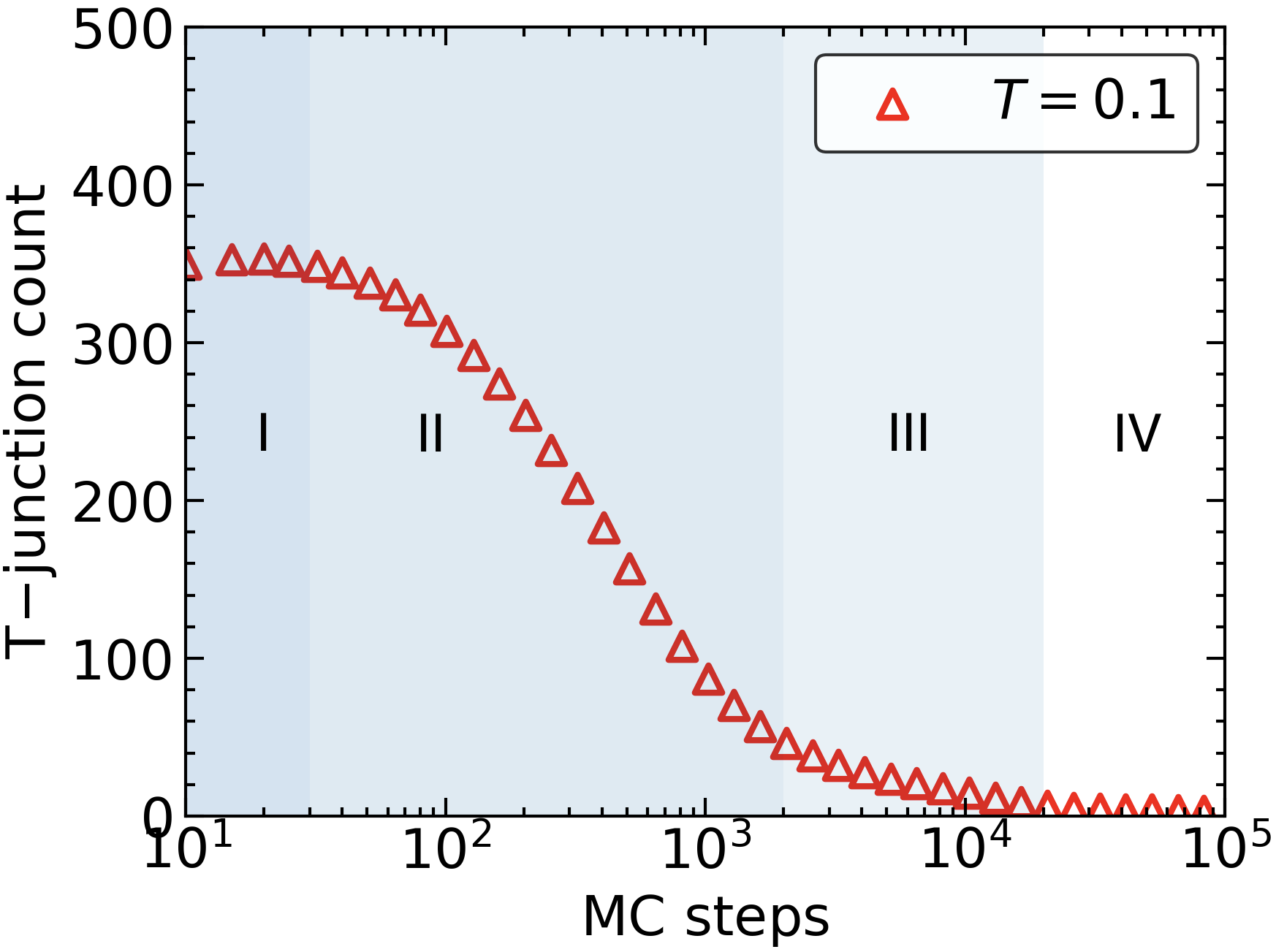}}

\subfloat[\label{fig:T-junctionCountTimeDependanceT0.3}]{\includegraphics[width=0.475\textwidth]{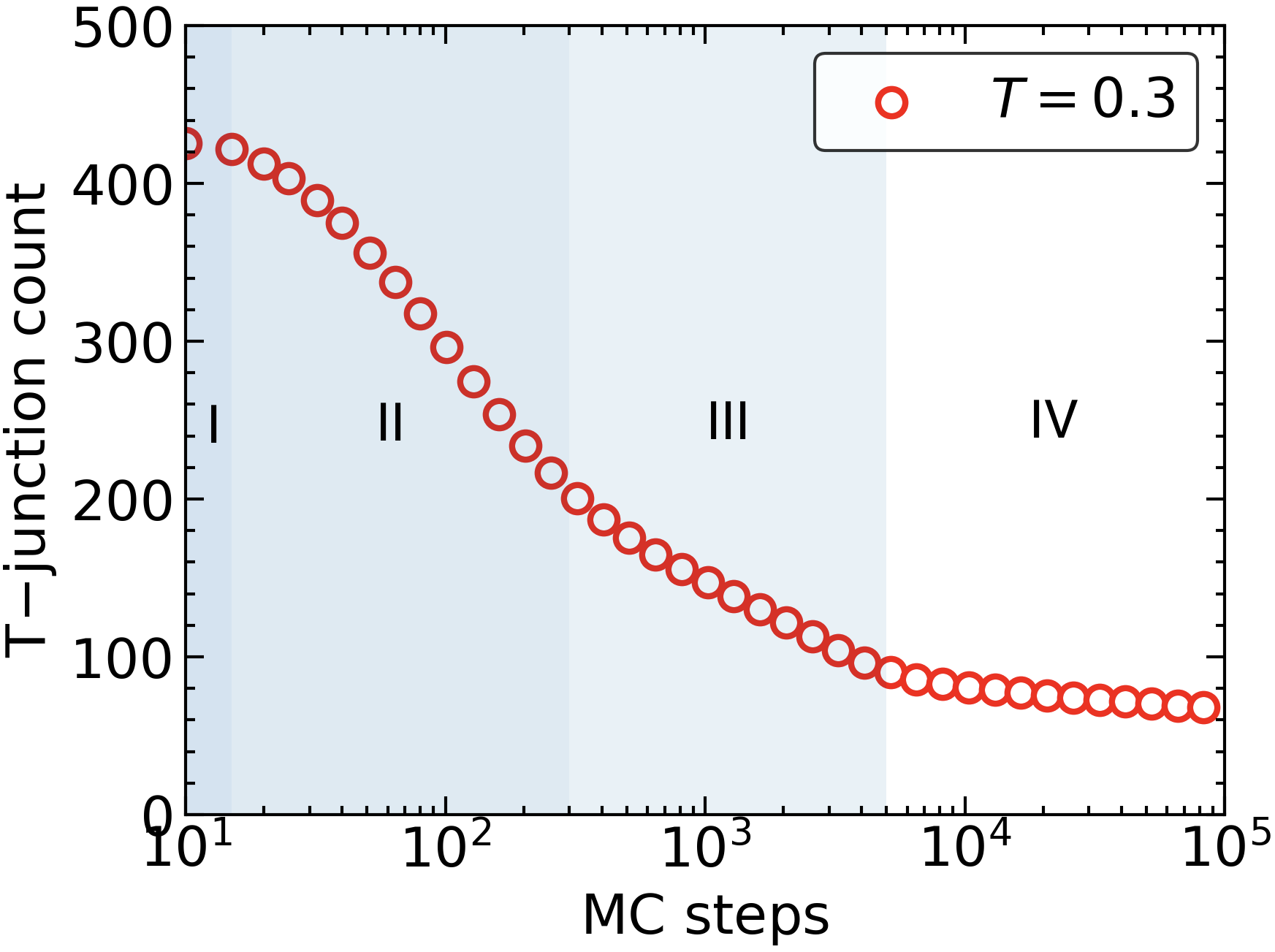}}

\caption{Evolution of the T-junction count with Monte Carlo steps for $L=100$ and $\delta=1.1$ at (a) $T=0.1$ and (b) $T=0.3$. The shaded regions, labelled I–IV, indicate four successive stages of the evolution. In regime I, the T-junction population increases as locally formed horizontal and vertical stripes organize into orientational super-domains. In regime II, the T-junction count decreases as the super-domains reorganize and grow through the motion and elimination of their boundary defects. Regime III corresponds to the single-super-domain stage, where one stripe orientation occupies most of the system, and the rate of T-junction elimination decreases as only a small number of residual boundaries remain. Regime IV represents the late-time equilibrium state. At $T=0.1$, the T-junction count approaches zero as a globally oriented stripe state is established, whereas at $T=0.3$ it approaches a finite equilibrium value due to thermally generated T-junctions.\label{fig:T-junctionCountTimeDependance}}

\end{figure}


Figure~\ref{fig:T-junctionCountTimeDependance} shows the T-junction count as a function of Monte Carlo steps for $L=100$ and $\delta=1.1$ at $T=0.1$ and $T=0.3$. The evolution can be separated into four dynamical regimes, indicated by the shaded regions.  In regime I, the shortest-time regime, the T-junction population increases as the initially disordered spin configuration develops local stripe order. Horizontal and vertical stripe segments form in different parts of the system and reorganize into respective super-domains.  As super-domains establish, boundaries form, leading to an increase in the number of T-junctions. This regime corresponds to the formation of the super-domain structure.  In regime II, once super-domains form, the T-junction populations begin to decrease as their boundaries start to move and reorganize. T-junctions associated with these boundaries are displaced and eliminated as super-domains grow, shrink, and merge. The reduction in defect populations reflects the main time of super-domain growth and orientational reorganization.

Next in time (10$^4$ to 10$^5$ MCS), the system enters regime III once it has largely reached a single-super-domain state with a single stripe orientation occupying most of the system.  In this regime, only a small number of residual boundaries with defects remain since fewer competing super-domains are available for elimination. The reduction of defects marks the crossover from the main super-domain growth regime to the final approach toward equilibrium.  Finally, at regime IV, the system reaches its equilibrium state.  For $T=0.1$, the remaining T-junctions from percolation super-domains are gradually eliminated, and the T-junction count approaches zero. For a higher temperature, $T=0.3$ (Fig.~\ref{fig:T-junctionCountTimeDependanceT0.3}), the T-junction count approaches a plateau since thermal fluctuations continually generate a small number of super-domains and their associated defects.


\begin{figure}[t!]

\centering

\includegraphics[width=0.475\textwidth]{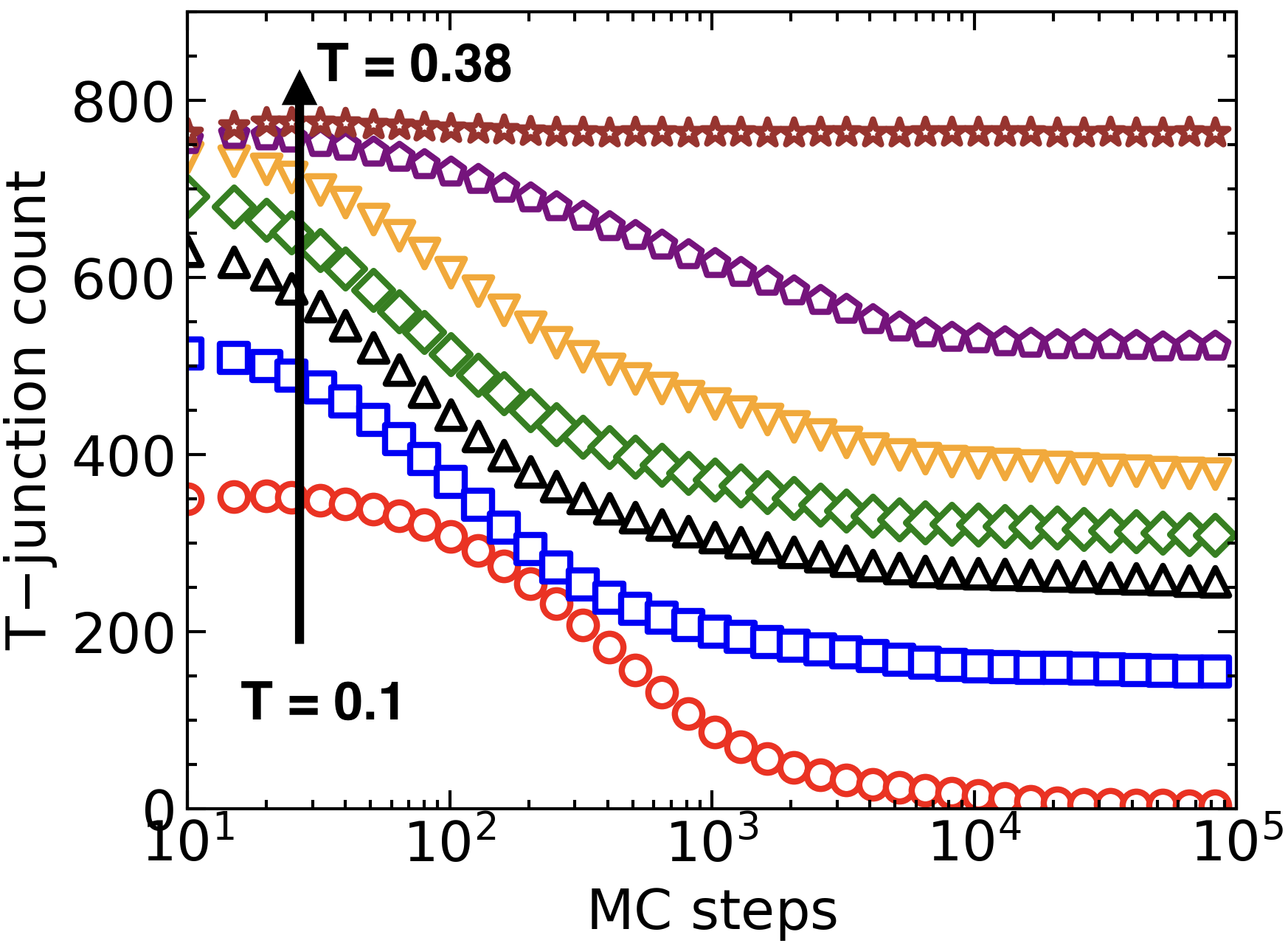}

\caption{T-junction count as a function of Monte Carlo steps for $L=100$ and $\delta=1.1$ at several quenching temperatures $(T = 0.1[\textcolor{red}{\pmb{\circ}}], 0.15[\textcolor{blue}{\pmb{\square}}], 0.2[\textcolor{black}{\pmb{\triangle}}], 0.25[\textcolor{green!50!black}{\pmb{\diamond}}], 0.3[\textcolor{orange}{\pmb{\triangledown}}], 0.35[\textcolor{purple}{\pmb{\pentagon}}], 0.38[\textcolor{red!60!black}{\pmb{\star}}] )$. For visual clarity, the curves are vertically offset by adding constants of $0$, $150$, $250$, $290$, $310$, $310$, and $310$ for $T=0.1$, $0.15$, $0.2$, $0.25$, $0.3$, $0.35$, and $0.38$, respectively.\label{fig:T-junctionCountTempDependance}}

\end{figure}


This difference between $T=0.1$ and $T=0.3$ T-junction evolution indicates that temperature affects the time scales of the dynamical regimes and the T-junction equilibrium. To examine this behaviour systematically, Fig.~\ref{fig:T-junctionCountTempDependance} presents the T-junction time-evolution for $L=100$ and $\delta=1.1$ for quench temperatures from $T=0.1$ to 0.38.  At low quench temperatures, the initial increase in the T-junctions happens over a relatively wide range of times, but with warming this regime becomes progressively shorter and the increase in T-junctions becomes smaller. The subsequent regimes of rapid T-junction elimination also shift to shorter times, indicating that the system progresses through the early stages of ``locking-in'' to a single super-domain more quickly with higher quench temperatures (i.e. closer to $T_C$), as expected.  More interesting is that the magnitude of this T-junction plateau increases with temperature.


\begin{figure}[b!]

\centering

\includegraphics[width=0.475\textwidth]{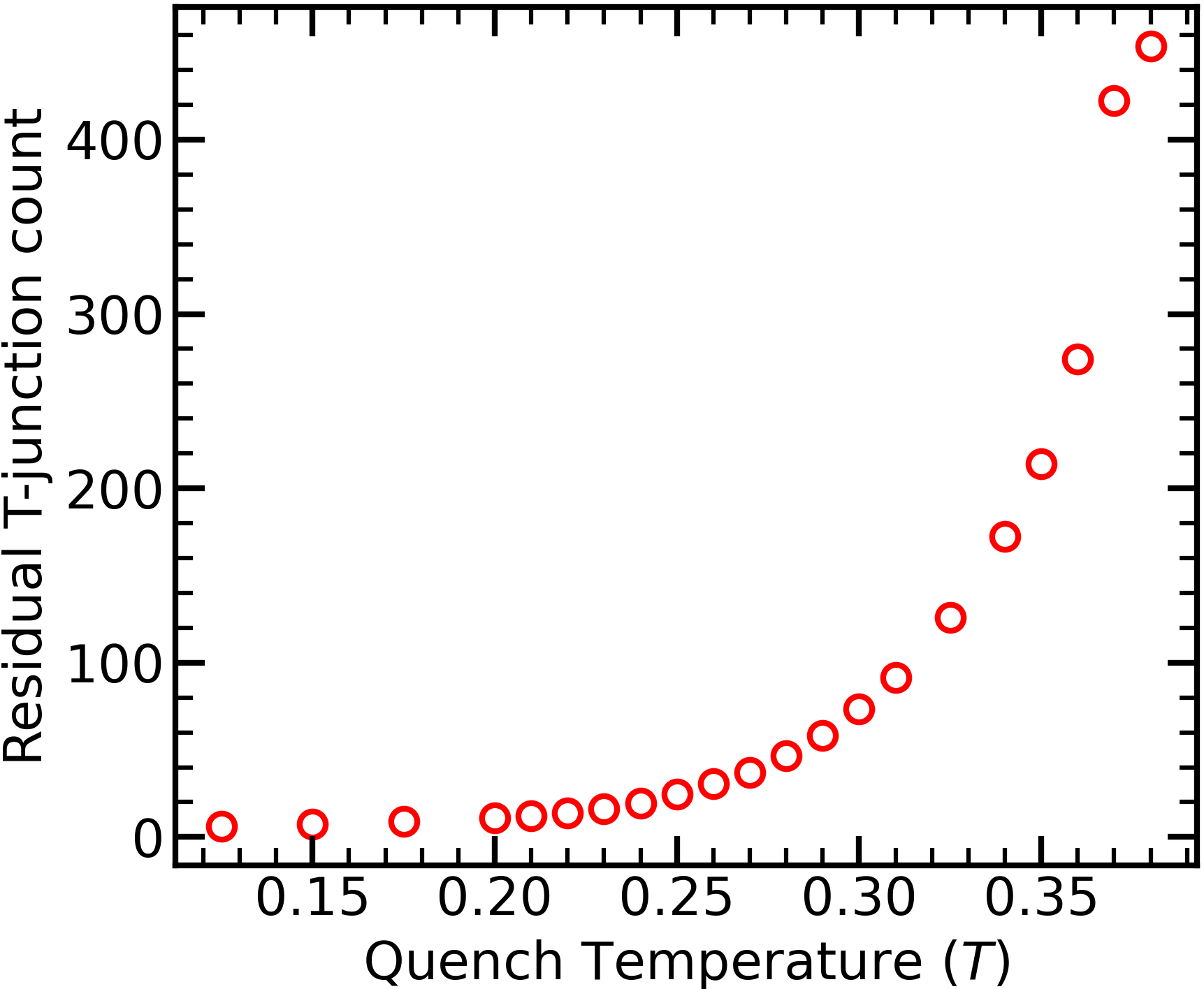}

\caption{Residual equilibrium T-junction count as a function of temperature for $L=100$ and $\delta=1.1$. The residual population remains small at low temperatures but increases rapidly as the temperature approaches the stripe-ordering transition, $T_C \simeq 0.38$.\label{fig:ResidualT-junctionCount}}

\end{figure}


The residual T-junctions at long times arise from thermal fluctuations near the equilibrium stripe state. These fluctuations locally disturb the ordered stripe pattern and create regions of the competing orientation, together with their associated T-junctions. T-junctions are therefore continuously created and eliminated near equilibrium, resulting in a nonzero time-averaged population. 

The temperature dependence of this residual population is shown in Fig.~\ref{fig:ResidualT-junctionCount}. At low temperatures, the equilibrium T-junction count remains small and changes only weakly with temperature. Above approximately $T\sim0.25$, however, the residual population begins to increase more rapidly, rising from a few tens of T-junctions to more than $400$ as the temperature approaches the stripe-ordering transition at $T_C\simeq0.38$. This rapid increase reflects the increasing strength of thermal fluctuations near $T_C$. Overall, increasing the quench temperature shortens the time required for super-domain evolution while substantially increasing the population of thermally generated T-junctions that remain at equilibrium.


\begin{figure}[t!]

\centering

\subfloat[\label{fig:T-junctionCountSizeDependenceT0.1}]{\includegraphics[width=0.475\textwidth]{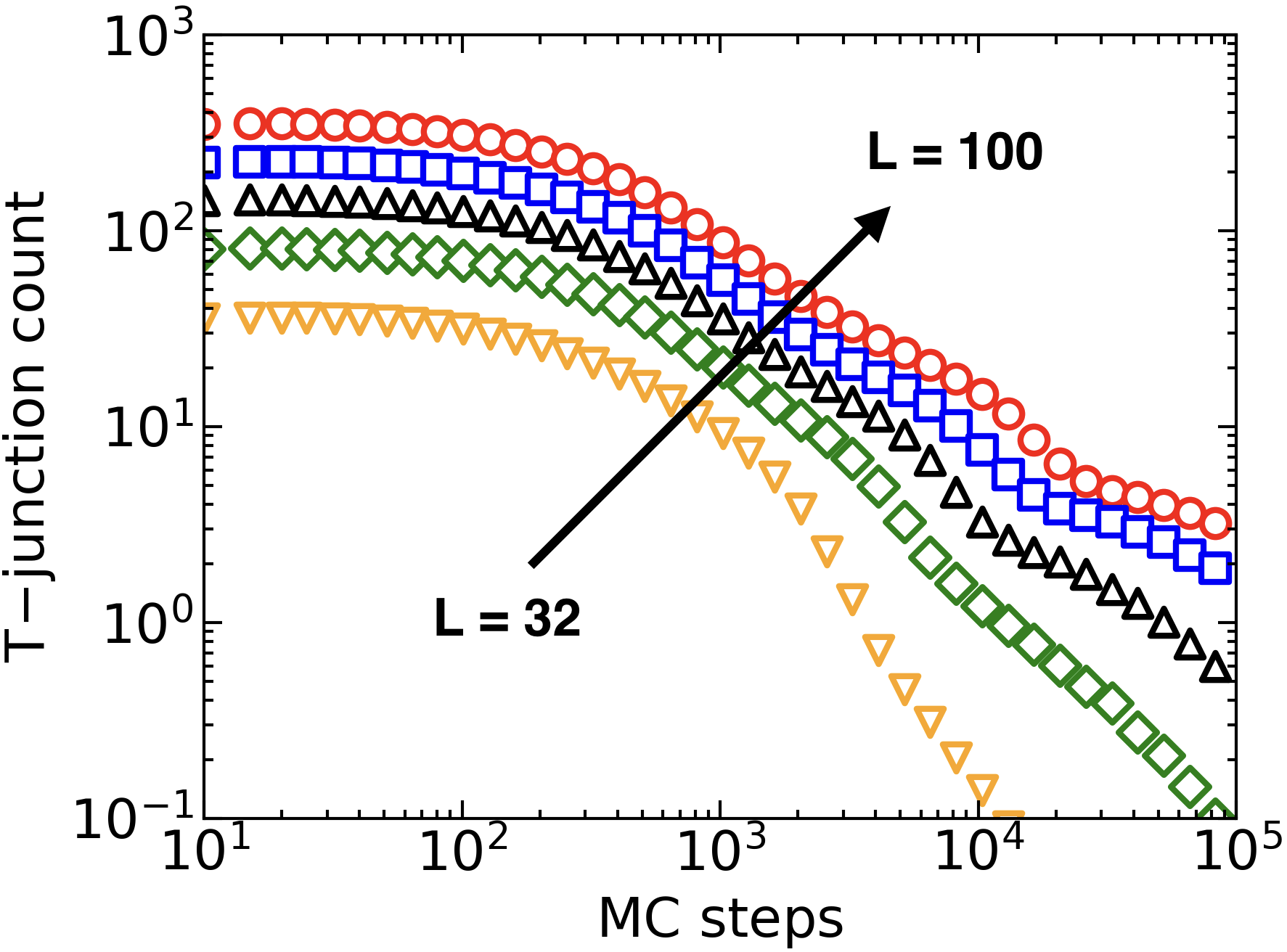}}

\subfloat[\label{fig:T-junctionCountSizeDependenceT0.3}]{\includegraphics[width=0.475\textwidth]{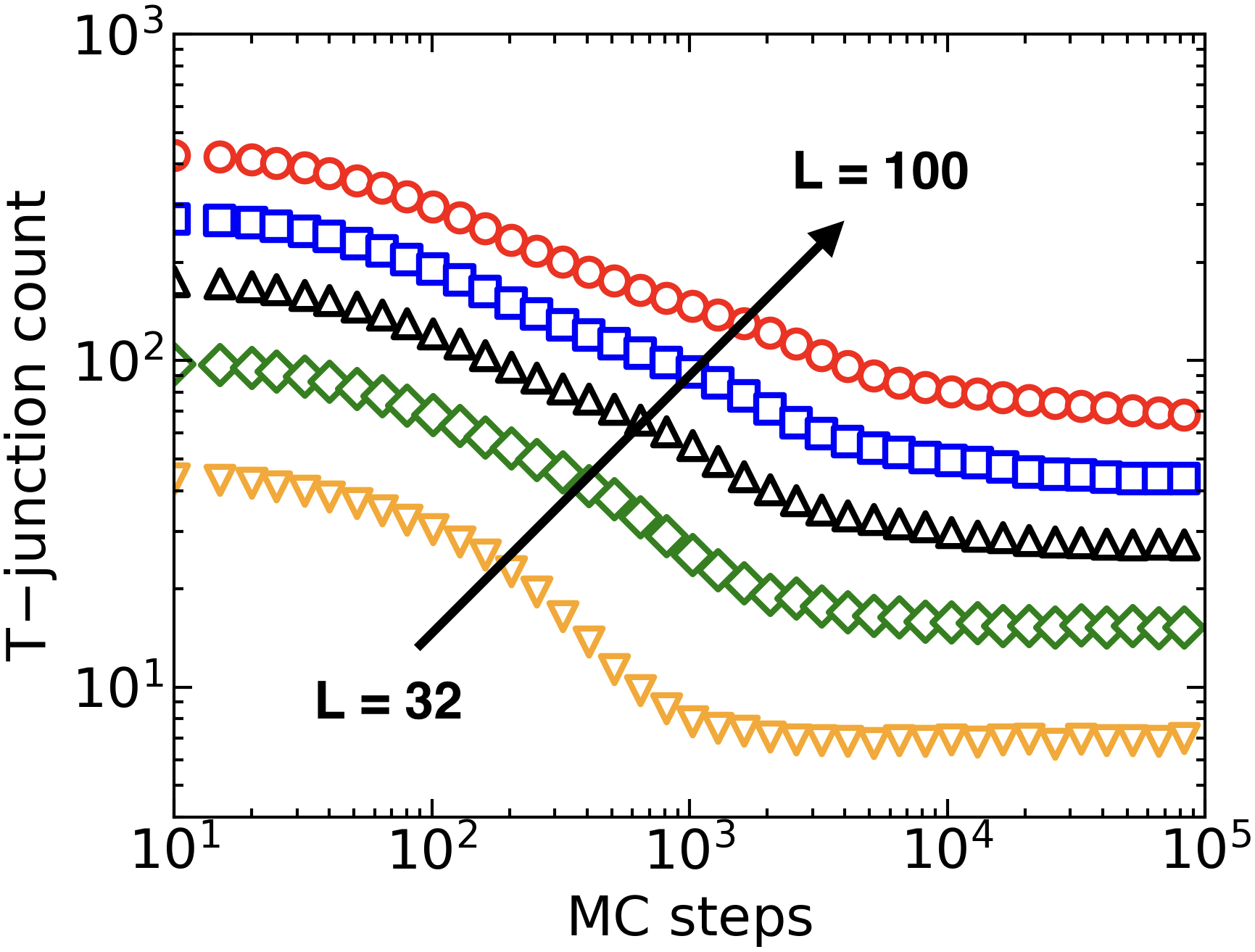}}

\caption{T-junction count as a function of Monte Carlo steps for system sizes $L = 100[\textcolor{red}{\pmb{\circ}}]$, $80[\textcolor{blue}{\pmb{\square}}]$, $64[\textcolor{black}{\pmb{\triangle}}]$, $48[\textcolor{green!50!black}{\pmb{\diamond}}]$, and $32[\textcolor{orange}{\pmb{\triangledown}}]$, at $\delta=1.1$, for (a) $T=0.1$ and (b) $T=0.3$. The dynamics exhibit distinct temporal regimes, with the crossover between regimes depending on system size: smaller systems enter the later-time regimes earlier, while larger systems require longer times. At $T=0.3$, the T-junction count eventually approaches a size-dependent finite plateau due to thermal fluctuations.\label{fig:T-junctionCountSizeDependence}}

\end{figure}


Are the times over these different evolution regimes intrinsic to the dynamics, or do they depend on the (finite) size of the system?  Figure~\ref{fig:T-junctionCountSizeDependence} shows the T-junctions temporal evolution for sizes between $L=32$ to $100$ with $\delta=1.1$ at $T=0.1$ and $T=0.3$.  A clear size dependence is observed in the crossover between the different stages. Smaller systems pass through the initial and intermediate regimes earlier and enter the late-time regime with fewer Monte Carlo steps.  As the system size increases, these crossovers shift to requiring more MCS to evolve and last for increasing amount.  These results reveal that larger systems require more time to eliminate boundaries between super-domains and to approach the final overall super-domain; a fully ordered (spin-aligned) configuration is never reached.  With more spins available, even near $T_C$, small domains always nucleate, grow, shrink and annihilate via T-junction defects.


\begin{figure}[b!]

\centering

\subfloat[\label{fig:T-junctionPairCorrelation_LeftRight}]{\includegraphics[width=0.475\textwidth]{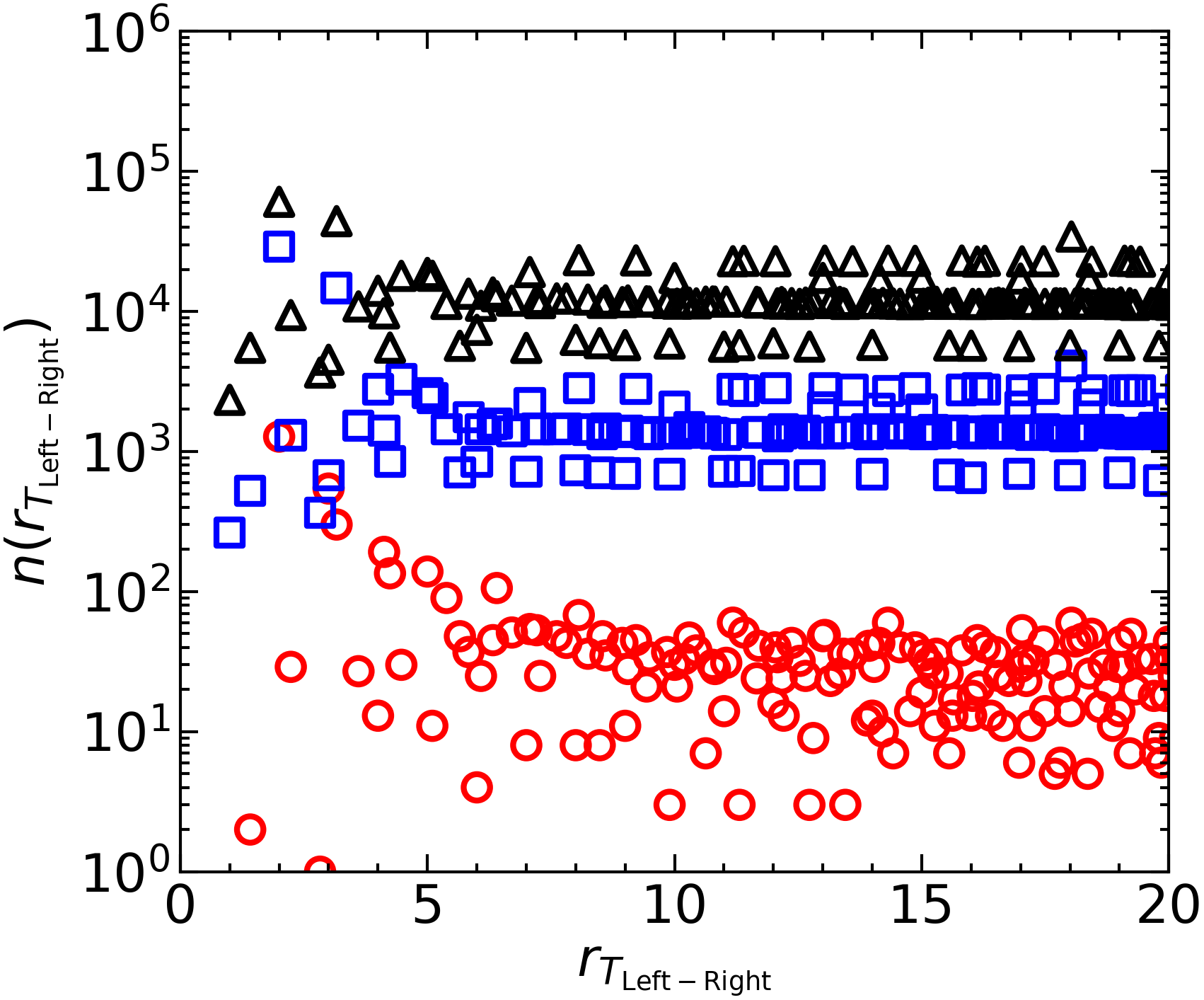}}

\subfloat[\label{fig:T-junctionPairCorrelation_UpDown}]{\includegraphics[width=0.475\textwidth]{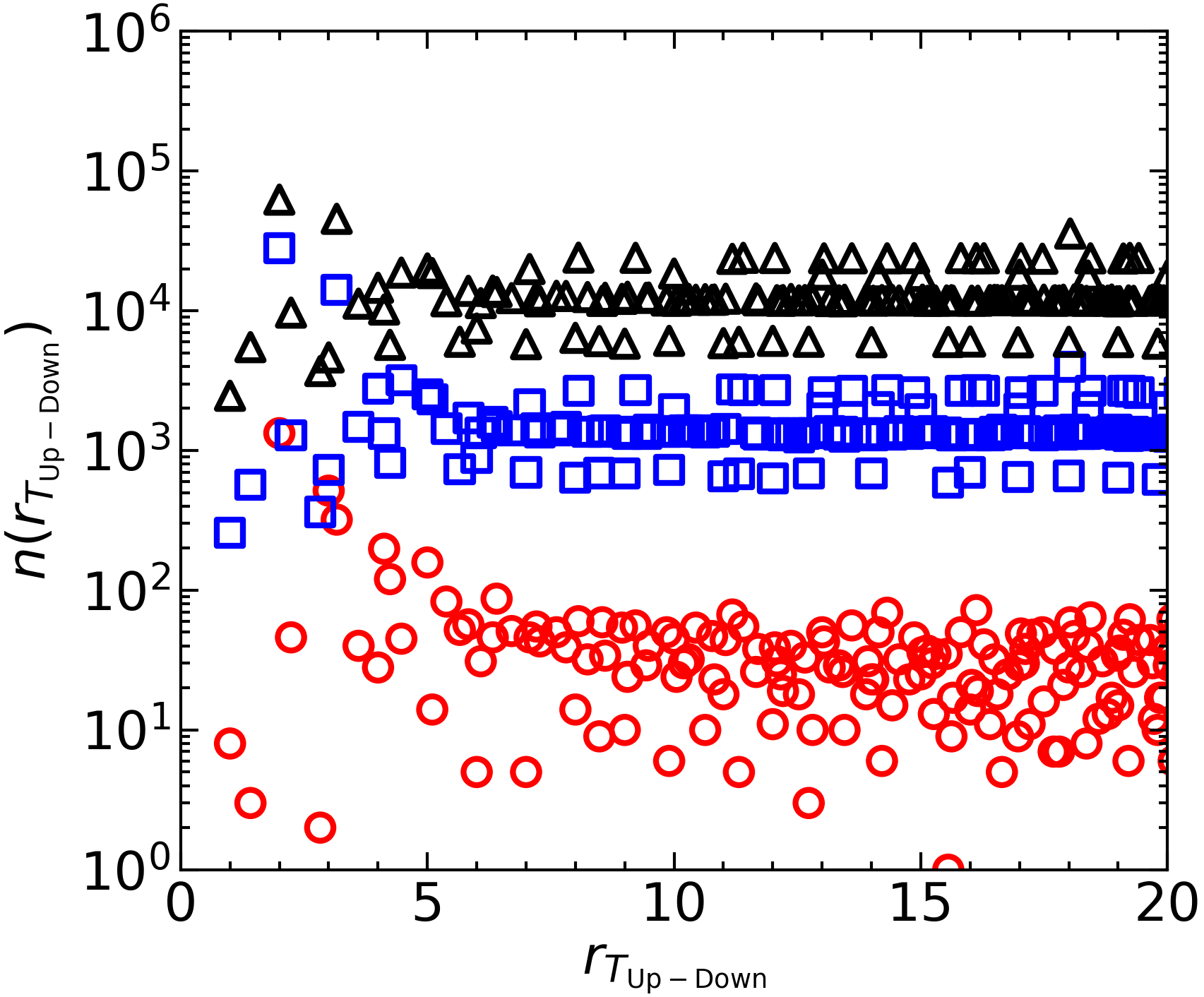}}

\caption{Pair-separation population of T-junctions for $L=100$, $\delta=1.1$, and $t=10^5$ Monte Carlo step at temperatures ($T= 0.2[\textcolor{red}{\pmb{\circ}}], 0.3[\textcolor{blue}{\pmb{\square}}], 0.35[\textcolor{black}{\pmb{\triangle}}]$), where the T-junction count saturates to a finite value even after the system has equilibrated. (a) Population $n(r_{T_{Left-Right}})$ of separations between left and right T-junctions. (b) Population $n(r_{T_{Up-Down}})$ of separations between up and down T-junctions. The increase in the population with temperature reflects the larger number of thermally generated T-junctions in the equilibrated stripe state, while the increasing spread toward larger separations indicates the formation of thermally generated super-domains of larger size.\label{fig:T-junctionPairCorrelation}}

\end{figure}


While the number of T-junctions measures the number of defects, it does not provide information about the spatial arrangement of strip domains and super-domains. To characterize these, we examine the population of separations between T-junction pairs in the equilibrium (plateau over large MCS) regime where a finite number of defects persists because of thermal fluctuations, as shown in Fig.~\ref{fig:T-junctionPairCorrelation}.  For a minority of super-domain embedded in a background of the opposite stripe orientation, T-junctions occur on two opposite sides of the super-domain. The distance between a corresponding pair provides a measure of the super-domain size, and we distinguish between \textbf{\emph{left--right}} and \textbf{\emph{up--down}} T-junction pairs, with the correlation $r$ between separation populations denoted by $n(r_{T_{\mathrm{Left-Right}}})$ and $n(r_{T_{\mathrm{Up-Down}}})$, respectively.

\begin{figure}[t!]

\centering

\includegraphics[width=0.475\textwidth]{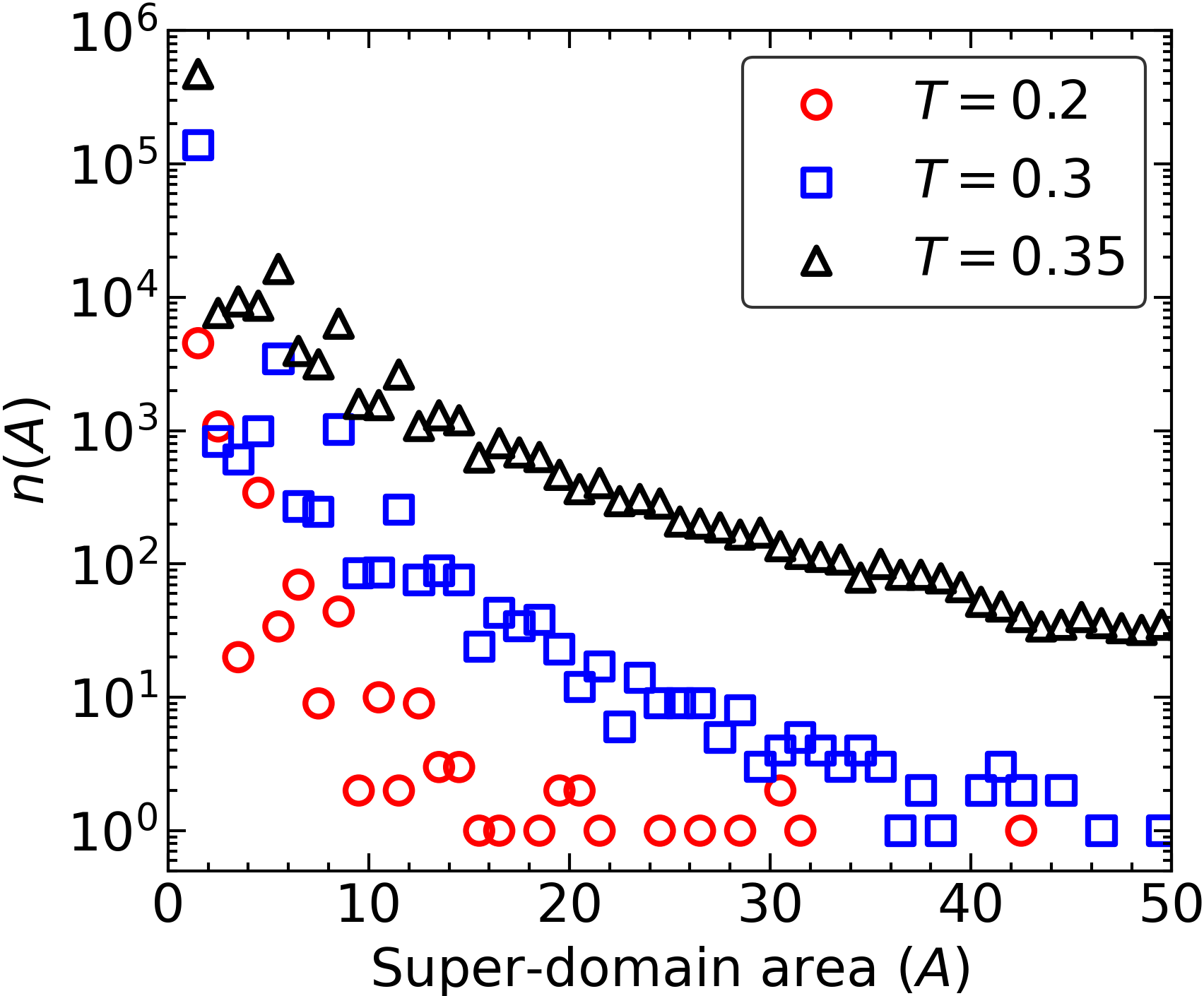}

\caption{Super-domain area population $n(A)$ for $L=100$, $\delta=1.1$, and $t=10^5$ Monte Carlo step at different temperatures  ($T= 0.2[\textcolor{red}{\pmb{\circ}}], 0.3[\textcolor{blue}{\pmb{\square}}], 0.35[\textcolor{black}{\pmb{\triangle}}]$). With increasing temperature, the population broadens and extends toward larger super-domain areas, indicating that thermal fluctuations generate both a larger number of super-domains and super-domains of larger size in the equilibrated stripe state. This observation is consistent with the T-junction pair-separation distributions, where the increasing spread toward larger separations with temperature indicates the presence of larger thermally generated super-domains.\label{fig:Super-domainAreaPopulation}}

\end{figure}

Figures \ref{fig:T-junctionPairCorrelation_LeftRight} and~\ref{fig:T-junctionPairCorrelation_UpDown} present the two separation populations for $L=100$ and $\delta=1.1$ at $T=0.2$, $0.3$, and $0.35$ at $10^5$ Monte Carlo step. The correlations between pairs identify thermally generated defects in equilibrated stripe-domains, and as the quench temperature approaches $T_C$, the T-junction correlations increase and extend toward larger separations. These trends are consistent with a higher equilibrium T-junction count near $T_C$, with T-junction pairs having larger separations as super-domains grow.  This is most clearly shown by examining Fig.~\ref{fig:Super-domainAreaPopulation} that exhibits the super-domain area population $n(A)$ for the same system parameters and quench temperatures. As the quench temperature increases, the number of super-domains increases, and the population extends toward larger $A$.


\begin{figure}[b!]

\centering

\subfloat[\label{fig:T-junctionCountTempDependanceStripeWidth2}]{\includegraphics[width=0.475\textwidth]{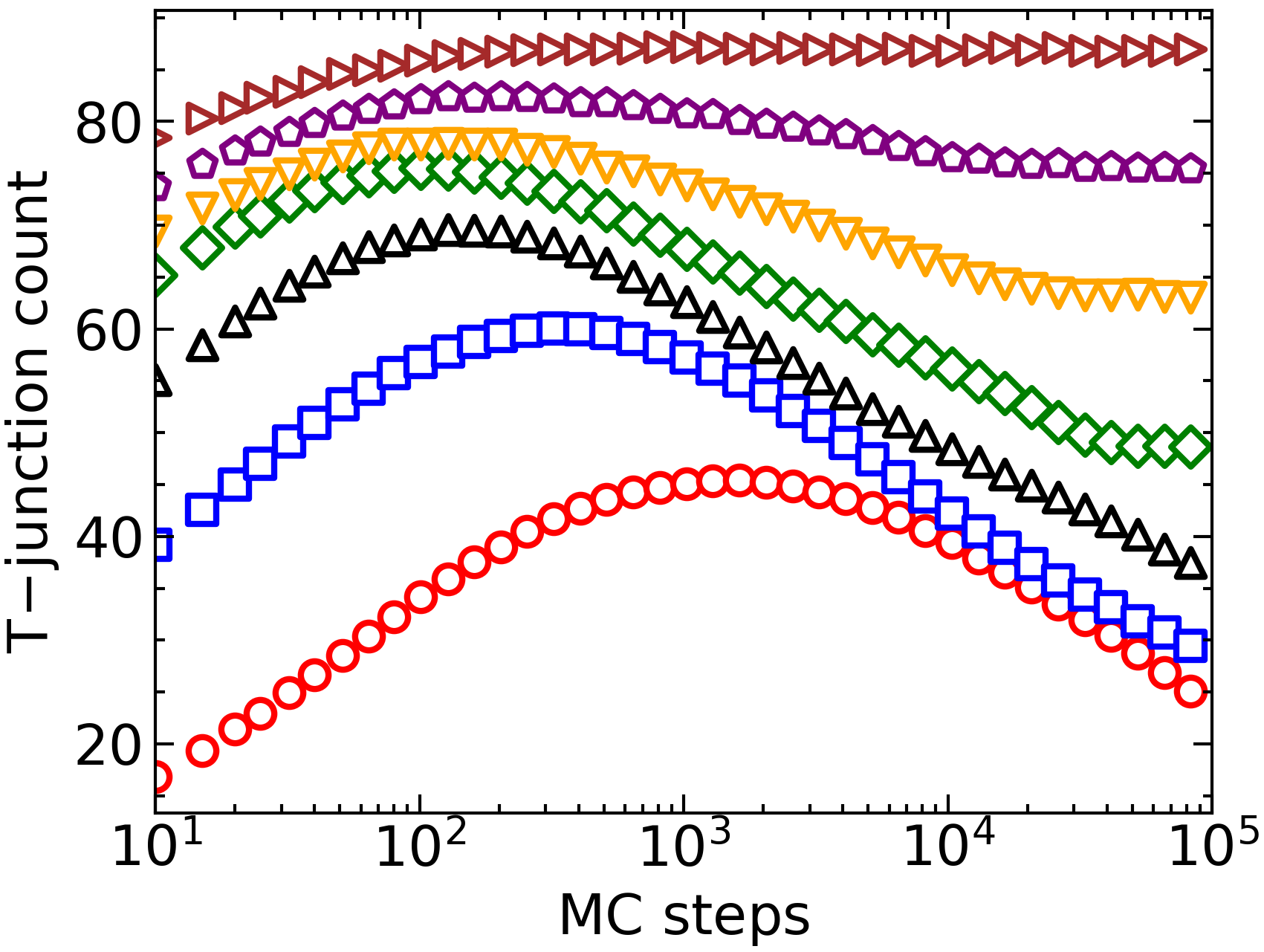}}

\subfloat[\label{fig:T-junctionCountTempDependanceStripeWidth3}]{\includegraphics[width=0.475\textwidth]{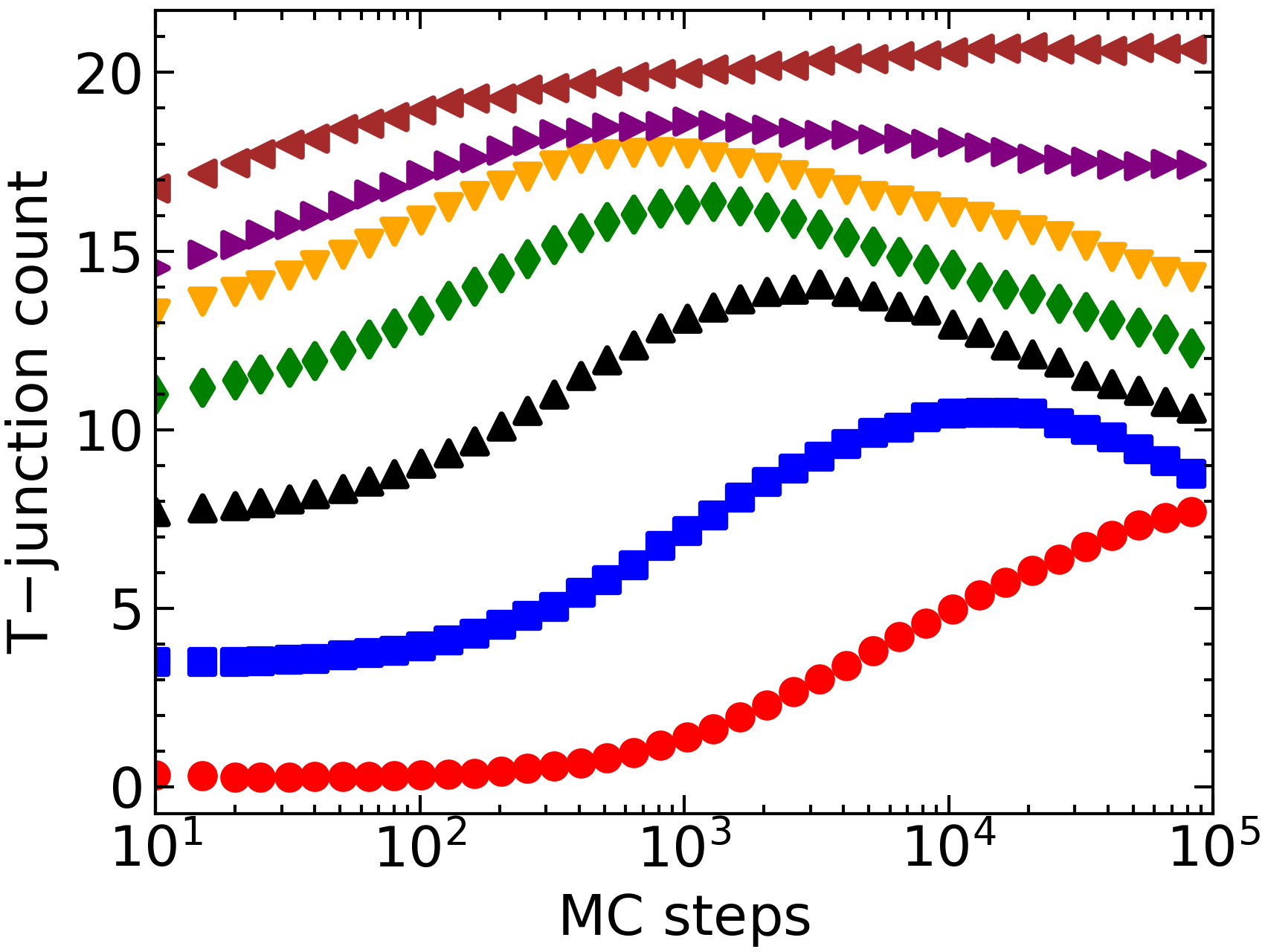}}

\caption{T-junction count as a function of Monte Carlo steps for systems of size $L=80$ with larger equilibrium stripe widths. (a) $\delta=2.0$, corresponding to stripe width $2$, at $T = 0.3[\textcolor{red}{\pmb{\circ}}], 0.4[\textcolor{blue}{\pmb{\square}}], 0.5[\textcolor{black}{\pmb{\triangle}}], 0.6[\textcolor{green!50!black}{\pmb{\diamond}}], 0.7[\textcolor{orange}{\pmb{\triangledown}}], 0.75[\textcolor{purple}{\pmb{\pentagon}}], 0.8[\textcolor{red!60!black}{\pmb{\triangleright}}]$. (b) $\delta=2.5$, corresponding to stripe width $3$, at $T = 0.4[\textcolor{red}{\bullet}], 0.5[\textcolor{blue}{\blacksquare}], 0.6[\textcolor{black}{\pmb{\blacktriangle}}], 0.7[\textcolor{green!50!black}{\pmb{\blacklozenge}}], 0.8[\textcolor{orange}{\pmb{\blacktriangledown}}], 0.9[\textcolor{purple}{\pmb{\blacktriangleright}}]$, and $1.06[\textcolor{red!60!black}{\pmb{\blacktriangleleft}}] $. As in the stripe-width-$1$ case, the T-junction population shows an initial increase followed by a reduction at later times, with the relative duration and magnitude of these regimes changing with temperature. For visual clarity, the curves in (a) are vertically offset by adding constants of $0$, $20$, $35$, $45$, $50$, $55$, and $60$, while those in (b) are offset by $0$, $4$, $7$, $9$, $10$, $12$, and $15$, in order of increasing temperature.\label{fig:T-junctionCountTempDependanceWiderStripes}}

\end{figure}


To examine whether the same behaviour persists for wider stripes, we consider $\delta=2.0$ and $\delta=2.5$, for which the equilibrium stripe widths are $2$ and $3$, respectively. Figure\ref{fig:T-junctionCountTempDependanceWiderStripes} shows the corresponding T-junction counts at several quenching temperatures. The overall evolution remains qualitatively similar to that observed for stripe width $1$. Following the quench, the T-junction population initially increases as the spins locally arrange into stripe segments and regions with different stripe orientations develop. Once these locally ordered regions have formed, the T-junction population begins to decrease as the defects move and are eliminated during the subsequent orientational ordering of the system. At sufficiently high temperatures, a finite population again persists at late times because thermal fluctuations continually generate local minority super-domains and the associated defects. Thus, the same sequence of dynamical processes identified for stripe width $1$ is also observed for stripe widths $2$ and $3$.


\begin{figure*}

\centering

\includegraphics[width=0.875\textwidth]{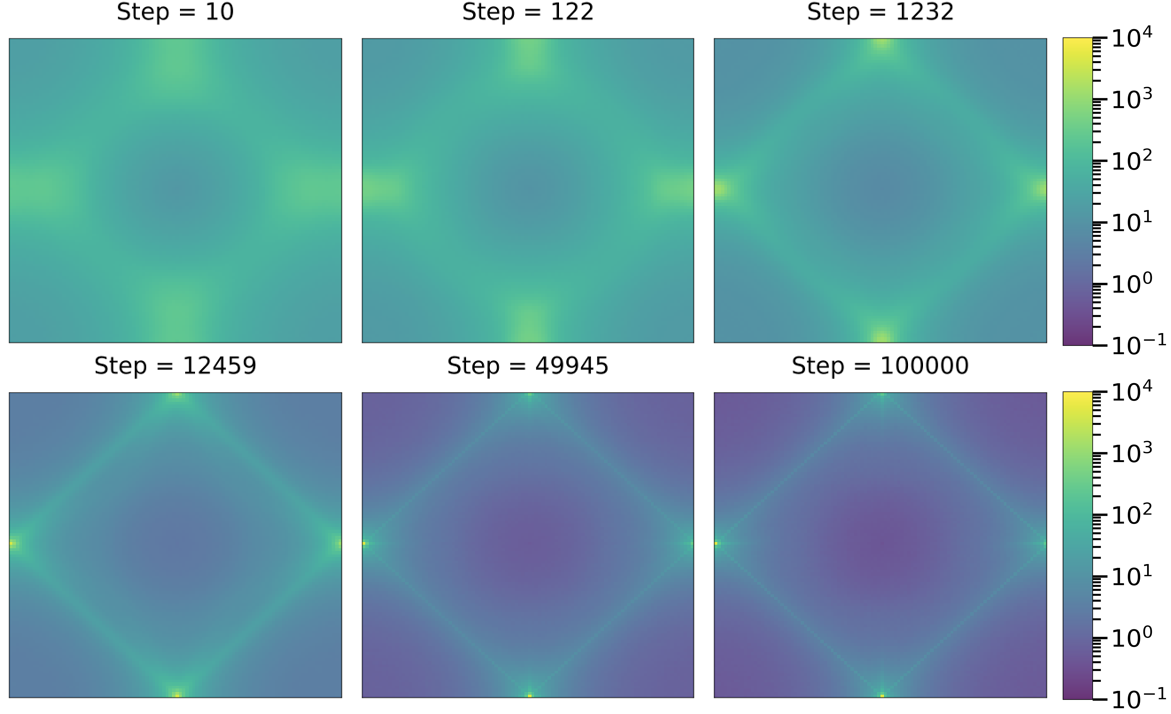}

\caption{Evolution of the structure factor during stripe ordering. The initially diffuse intensity sharpens into distinct peaks as Monte Carlo time increases, indicating the development of long-range stripe order and residual defect-related features.\label{fig:FFTEvolutionPathway}}

\end{figure*}


The main difference is the time (number of MCS) over which these processes occur. As the stripe width increases, a larger number of spins must organize collectively to establish the local stripe structure. Consequently, the formation of well-defined locally ordered stripe regions requires more Monte Carlo steps. This is reflected directly in the T-junction count: the initial regime in which the number of T-junctions increases extends over progressively longer times for larger stripe widths. The maximum in the T-junction population, which marks the crossover from the formation of domains to the subsequent elimination of their boundaries as super-domains form, shifts further on into the equilibration process with increasing stripe width.  These results indicate that increasing the stripe width does not change the underlying defect-mediated ordering mechanism. Instead, stripe width changes the evolutionary time periods primarily.


\subsection{Reciprocal-space signatures of defect evolution}

The preceding analysis describes the ordering process entirely in real space through the evolution of super-domains and their associated defects. An important question is whether the same microscopic evolution can also be identified in reciprocal space, where stripe ordering is accessible through scattering measurements\cite{thompson2008imaging, gupta2025anisotropic, PhysRevB.51.16033, indergand2023domain, singh2021late}. We therefore examine how the defect-mediated evolution described above is reflected in the structure factor.

Figure~\ref{fig:FFTEvolutionPathway} shows the evolution of the structure factor during the ordering process. At early times, the intensity is broadly distributed, reflecting the absence of well-developed stripe order. As the system evolves, the intensity becomes increasingly concentrated around the characteristic wave vectors of the stripe phase. At later times, these features become sharper as the system approaches a more ordered stripe configuration.

The structure factor's evolution in time (MCS) contains distinct signatures of the horizontal and vertical stripe populations. Horizontal stripes are periodic along the vertical direction in real space and therefore produce intensity maxima along the vertical reciprocal-space axis. Likewise, vertical stripes are periodic along the horizontal direction and generate intensity maxima along the horizontal reciprocal-space axis. The intensity of these peaks is related to the total area occupied by stripes of the corresponding orientation, whereas their widths reflect the spatial coherence of that stripe order.  When an orientational region is fragmented into many smaller pieces, the corresponding intensity becomes more diffuse. As larger and more coherent stripe regions develop, the intensities sharpen.  In addition to the principal stripe features, the structure factors contain diagonal features that originate from the geometry of the super-domain boundaries. As discussed earlier, the boundaries between horizontal and vertical super-domains are predominantly oriented along the $45^\circ$ and $135^\circ$ directions, producing the diagonal lines in reciprocal space. The width of these lines carries information about the spatial organization of the boundaries. When the super-domains are small and fragmented, the corners forming their boundaries are distributed over short segments and their alignment along the two diagonal directions is less coherent. The corresponding diagonal intensity is therefore broad. As the super-domains grow, longer boundary segments develop, and the corners become more coherently aligned along the $45^\circ$ and $135^\circ$ directions. This increased coherence is reflected by a progressive resolving of diagonal features.  The remaining diffuse intensity is from corners and T-junctions that locally disrupt the otherwise periodic stripe pattern and contribute intensity over a broad range of wave vectors. Away from the horizontal and vertical stripe intensities and the diagonal features, the background intensity can hold information about the defect population. To explain, when compared to earlier times or lower quench temperatures, since measured intensities are not necessarily absolute unless performed in transmission and a one-to-one correspondence of measured photon-to-spin in the system is possible. For example, a configuration containing a larger number of such defects would produce a stronger diffuse background intensity, whereas this contribution should reduce as defects are removed during the evolution.


\begin{figure*}

\centering

\subfloat[\label{fig:DefectAnhiliationRateT0.1}]{\includegraphics[width=0.475\textwidth]{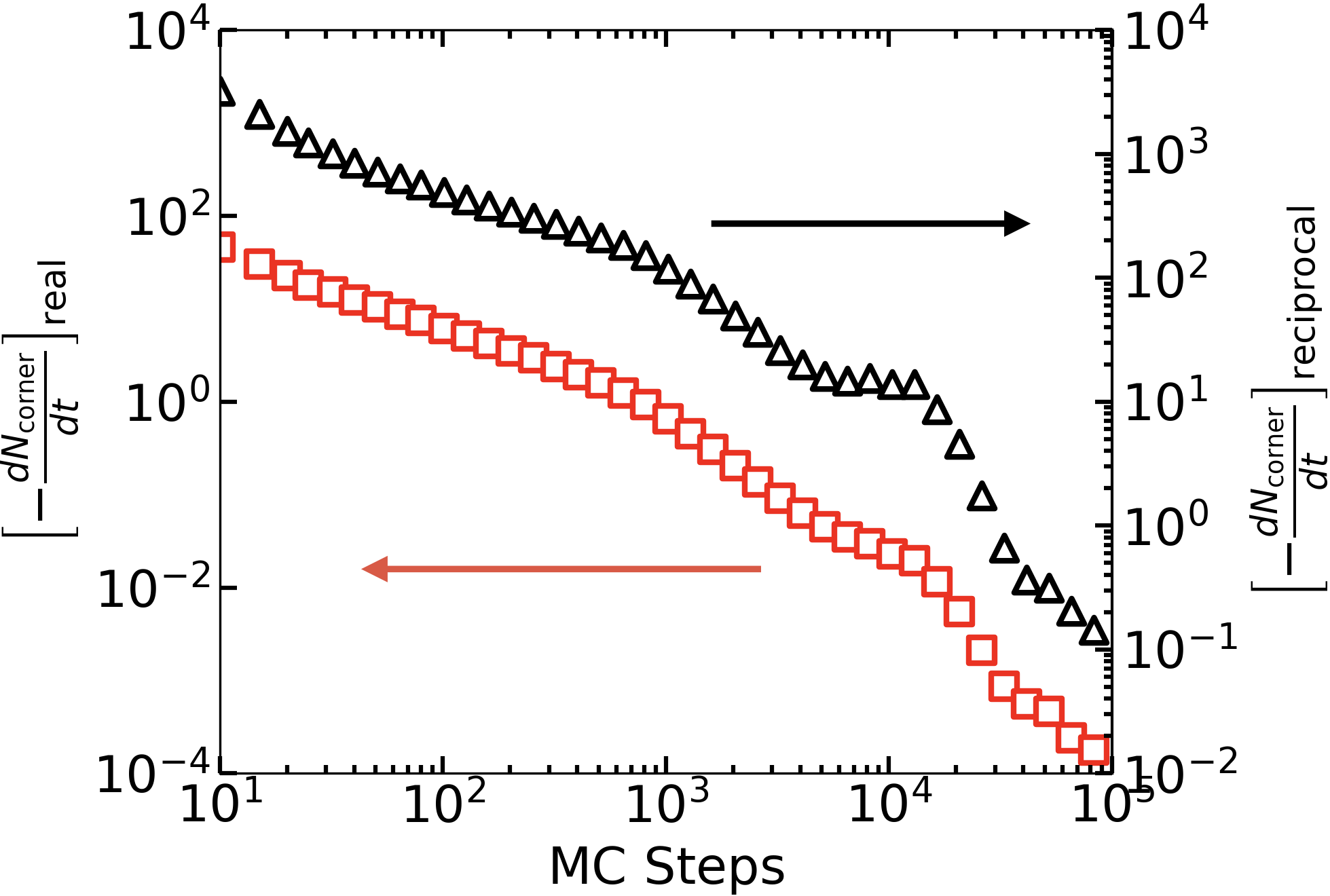}}
\hfill
\subfloat[\label{fig:DefectAnhiliationRateT0.2}]{\includegraphics[width=0.475\textwidth]{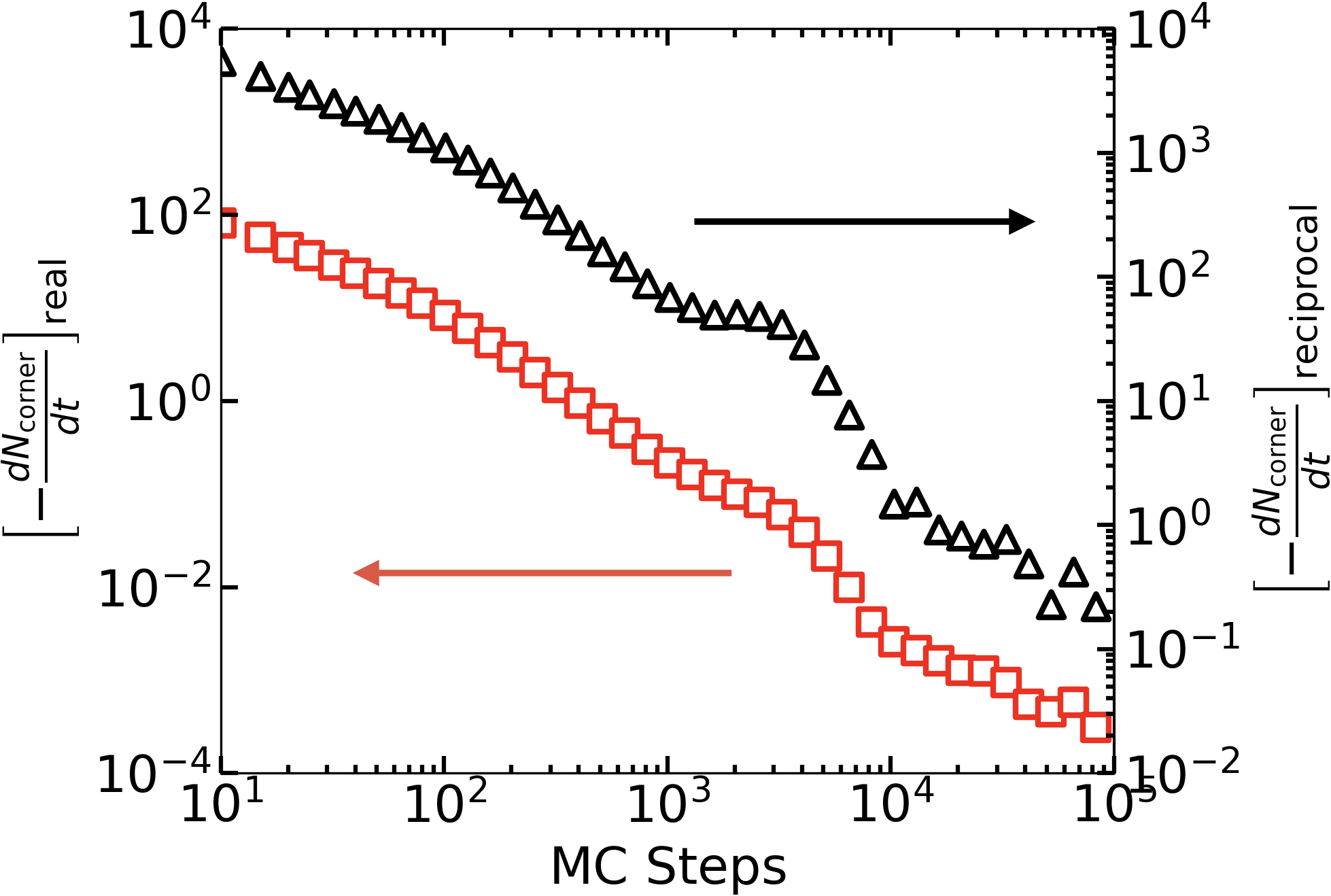}}

\subfloat[\label{fig:DefectAnhiliationRateT0.25}]{\includegraphics[width=0.475\textwidth]{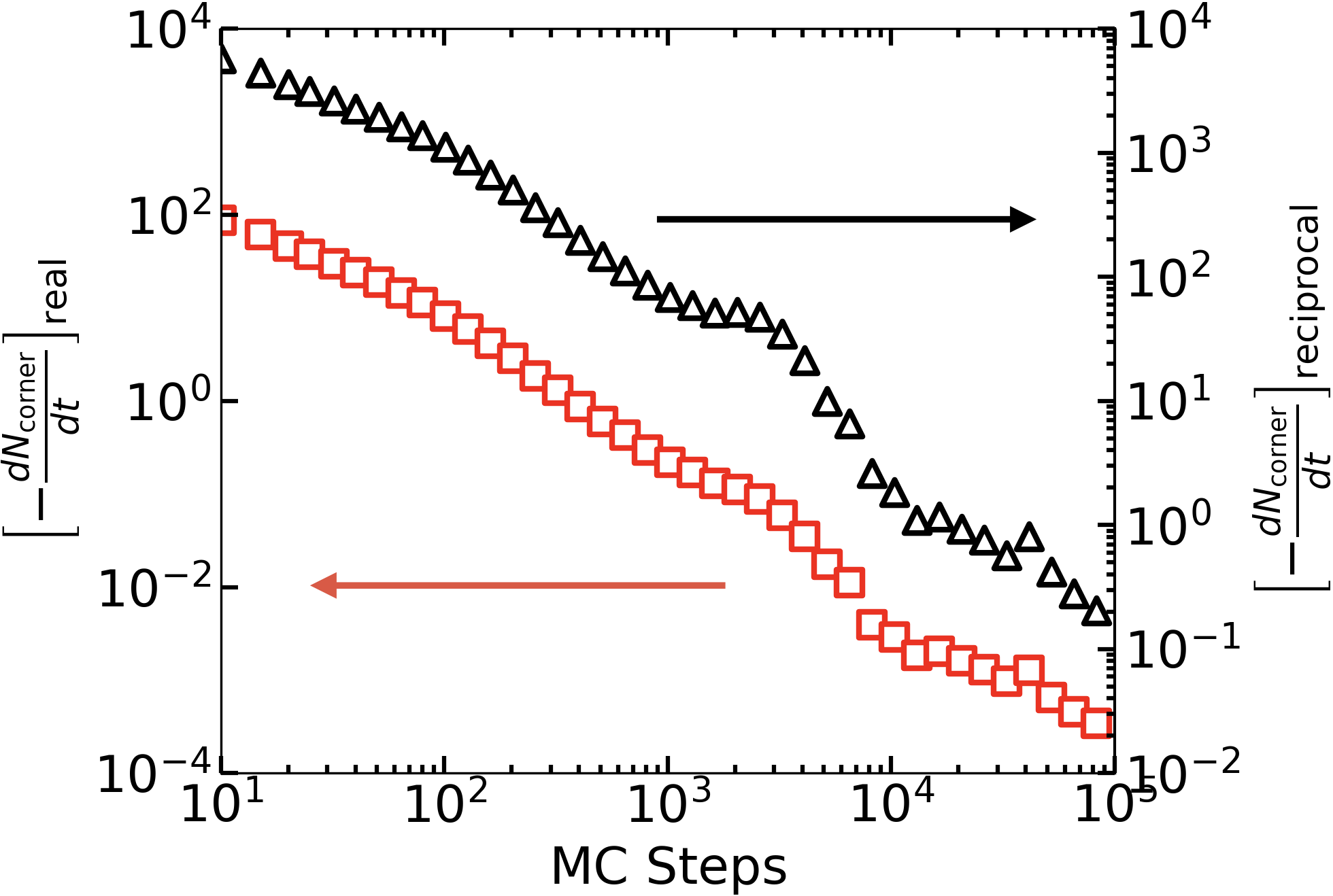}}
\hfill
\subfloat[\label{fig:DefectAnhiliationRateT0.3}]{\includegraphics[width=0.475\textwidth]{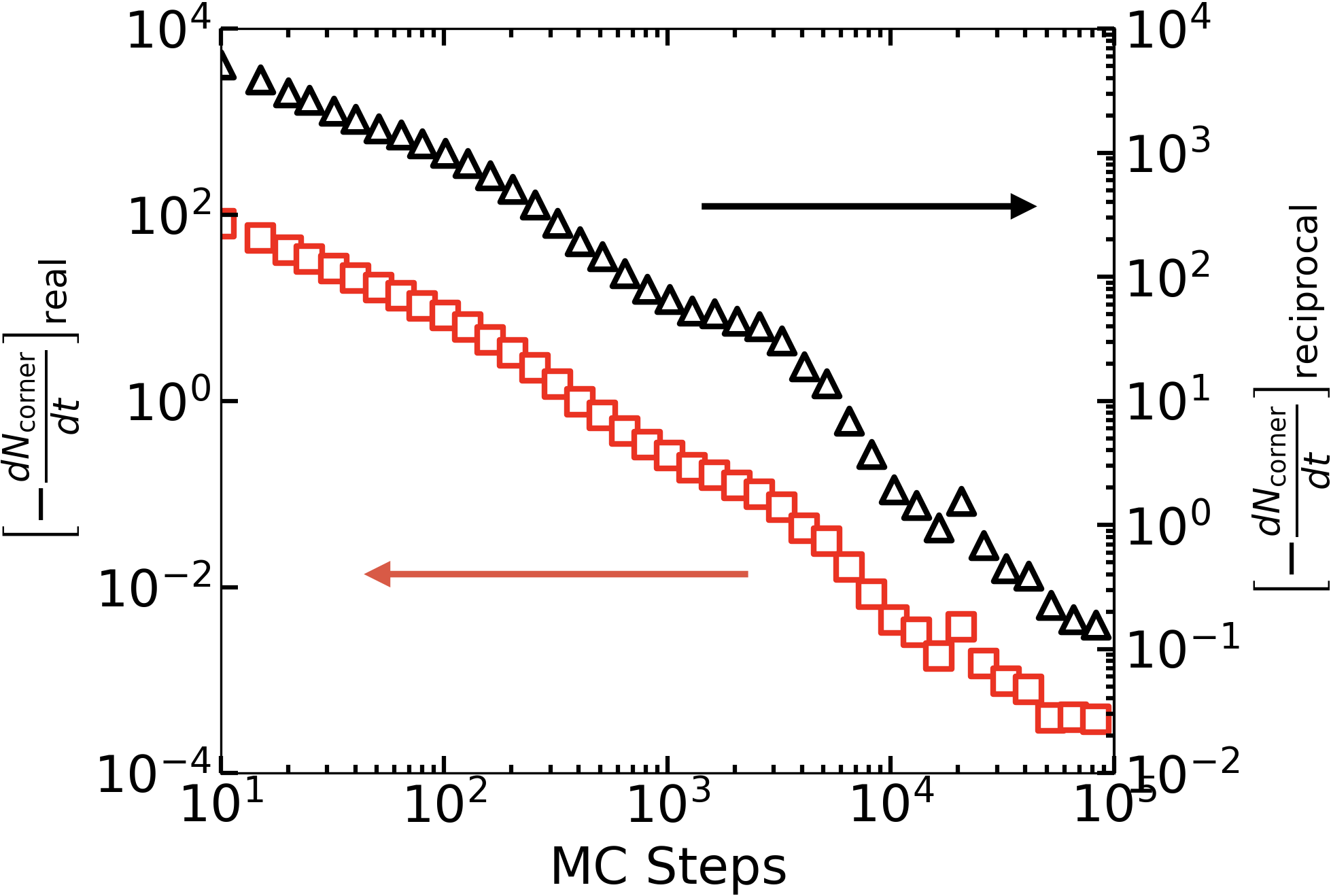}}

\caption{Rate of defect elimination measured independently in real space [$\textcolor{red}{\pmb{\square}}$] and reciprocal space [$\textcolor{black}{\pmb{\triangle}}$] for $L=100$ and $\delta=1.1$ at (a) $T=0.1$, (b) $T=0.2$, (c) $T=0.25$, and (d) $T=0.3$. The real-space quantity is obtained from the decay rate of the corner count, while the reciprocal-space quantity is obtained from the rate of change of the total structure-factor intensity. Although the two measures have different absolute magnitudes, they show the same overall decrease with Monte Carlo time and similar changes in dynamical behaviour. This correspondence indicates that the defect-elimination dynamics observed directly in real space are also reflected in reciprocal space.\label{fig:DefectAnhiliationRate}}

\end{figure*}


The structure-factor signatures discussed above suggest that the evolution of the defect populations can be tracked without identifying individual defects.  To test this, we compare the defect-elimination rates obtained from the defect counts (corners $+$ T-junctions $+$ I-junctions) with a corresponding rate extracted from the structure factor changes.  Figure~\ref{fig:DefectAnhiliationRate} presents the results for $L=100$ and $\delta=1.1$ at quenches to $T$=0.1, 0.2, 0.25, and 0.3. In all cases, the reciprocal-space rates follow the same overall evolution as the real-space rates.  Both quantities decrease longer into the simulation (MCS), and changes in their respective slopes occur at similar regimes as identified above. Correspondence between real-space and scattering ($k$-space) is therefore not limited to just near-equilibrium behaviours, but translates to changes in the dynamics of the system through the identified different stages of stripe reorganizations.  This can be understood from the way defects modify the stripe pattern. At early and intermediate times, the system contains a large number of super-domain boundaries, corners, and other local irregularities. These structures disrupt the periodicity of the stripes and contribute broadly to the reciprocal-space intensity. As the super-domains reorganize and their boundaries are eliminated, the stripe patterns become progressively more regular. The reduction in the real-space defect population is therefore accompanied by a corresponding change in the structure-factor intensity.


\section{Summary and conclusion}

We have investigated the nonequilibrium evolution of stripe-forming systems following a quench from a disordered state into the stripe-ordered regime. Using a two-dimensional Ising model with competing short-range exchange and long-range dipolar interactions, we focused on the microscopic process by which a system containing two equivalent stripe orientations develops global orientational order. Rather than describing this evolution only through a global order parameter, we followed the real-space structures that appear as the stripe pattern develops and reorganizes.

Following the quench, horizontal and vertical stripe segments first form locally and organize into larger orientationally ordered regions, which we identify as super-domains. The boundaries separating super-domains of different orientations contain characteristic defects, including corners, T-junctions, and I-junctions. We find that the subsequent evolution of the stripe pattern is controlled by the motion and elimination of these boundary defects. In particular, T-junction motion is directly associated with changes in super-domain size: vertical T-junctions mediate the shrinkage of vertical super-domains, whereas horizontal T-junctions mediate the shrinkage of horizontal super-domains. The imbalance between horizontal and vertical T-junction populations is correspondingly related to the stripe orientation that eventually becomes dominant. This establishes T-junctions as a microscopic indicator of the pathway through which competing stripe orientations are removed.

The time dependence of the total T-junction population provides a simple description of the different stages of this evolution. At early times, the T-junction population increases as local stripe order develops and differently oriented super-domains are formed. This is followed by a regime of defect elimination, during which T-junctions move and are eliminated as super-domains of the competing orientation shrink. At late times, the evolution slows as the system approaches equilibrium. At sufficiently low temperatures, the T-junction population approaches zero as a globally oriented stripe state is established. At higher temperatures, however, a finite equilibrium population remains because thermal fluctuations continually generate local minority super-domains and their associated defects.

Temperature changes both the time scales of this evolution and the nature of the late-time state. Increasing temperature shortens the early stages associated with local stripe formation and subsequent defect elimination, while increasing the equilibrium T-junction population. The spatial organization of these thermally generated defects provides further information about the structures present at equilibrium. The population of T-junction pair separations extends toward larger separations with increasing temperature, indicating that thermally generated minority super-domains can span larger linear dimensions. The corresponding super-domain area population likewise extends toward larger areas. These two measurements therefore provide complementary descriptions of the same thermal structures and connect the residual equilibrium defect population to the size of the minority super-domains that generate it.

The time scales also depend on the system size and on the equilibrium stripe width. Larger systems require longer times to eliminate orientational boundaries and reach the late-time regime, reflecting the larger spatial scale over which global orientational order must develop. Increasing the stripe width produces a different but related slowing of the early dynamics. For stripe widths two ($h=2$) and three ($h=3$), the same sequence of T-junction formation and elimination observed for stripe width one ($h=1$) remains present, showing that the underlying mechanism is unchanged. However, wider stripes require more Monte Carlo time to establish the local stripe structure. As a result, the initial regime in which super-domains and T-junctions are formed extends to progressively later times as the stripe width increases. The defect-mediated picture of orientational ordering is therefore not restricted to a particular stripe width; the principal change is in the characteristic time scales of the process.

Finally, we showed that the same evolution can be identified in reciprocal space. The structure factor contains distinct signatures of the stripe orientations, the diagonal boundaries between super-domains, and the defects that locally disrupt the periodic stripe structure. As the system evolves, the principal stripe peaks sharpen, the diagonal features associated with super-domain boundaries become more coherent, and the diffuse defect-related intensity decreases. Most importantly, the rate of change of the defect-related structure-factor intensity follows the same overall time dependence as the rate of defect removal measured directly in real space. The agreement between these independently determined quantities demonstrates that the microscopic defect-elimination process leaves a measurable signature in reciprocal space.

Taken together, these results provide a microscopic description of how global stripe orientation emerges after a quench. The system does not select a single orientation through a uniform collective rotation or a direct growth of one stripe orientation from the outset. Instead, local stripe order first develops independently in both equivalent orientations, producing orientational super-domains. Global order then emerges through the motion, interaction, and elimination of the defects separating these regions, with T-junctions playing a central role in their reorganization. Temperature, system size, and stripe width modify the time and length scales of this process without changing its underlying mechanism within the square-lattice model considered here. In systems with a different lattice symmetry, more than two equivalent stripe orientations, or an approximately continuous orientational degeneracy, the detailed junction structure and the relationship between junction orientation and super-domain shrinkage need not be the same. 

Nevertheless, the broader mechanism identified here may extend to systems in which the allowed stripe orientations are restricted to a small number of symmetry-related variants: orientational order can then develop through the motion and elimination of junctions separating competing stripe regions. Similar connections between domain evolution and defect dynamics have been identified in systems ranging from liquid crystals and active matter to lamellar polymers and other symmetry-broken phases\cite{huh2024universality, fumeron2023introduction, angheluta2026full, balch2023spatially, singh2021late, liang2026phase, indergand2023domain, pinna2025mechanisms, patel2022rapid, leniart2022pathway, blagojevic2023multiscale, mondal2024ordering}. Although the relevant defects and order-parameter symmetries differ between these systems, these studies provide a broader context in which domain growth can be understood through the microscopic structures associated with domain boundaries.

The connection between real-space defect dynamics and reciprocal-space intensity also provides a route toward experimental observation of this evolution. In systems where direct microscopic imaging of individual defects is difficult, time-resolved scattering measurements could provide access to the same defect-mediated dynamics through the evolution of the structure factor. The present results therefore establish a connection between the microscopic rearrangement of stripe-domain structures and experimentally accessible reciprocal-space measurements, providing a framework for studying nonequilibrium orientational ordering in stripe-forming systems.


\vspace*{12pt}

\begin{acknowledgments}
J.v.L. and R.L.S. acknowledge funding from the Natural Sciences and Engineering Research Council of Canada (RGPIN 2024-04882 and RGPIN 05011-18). Computational resources were enabled in part by support provided by the University of Manitoba Grex High Performance Computing Centre (RRID:SCR\_026342), the University of Manitoba (umanitoba.ca), and the Digital Research Alliance of Canada (alliancecan.ca).
\end{acknowledgments}

\bibliography{2D_stripes_1.bib}

@article{gleiser2003slow,
  title={Slow dynamics in a two-dimensional Ising model with competing interactions},
  author={Gleiser, Pablo M and Tamarit, Francisco A and Cannas, Sergio A and Montemurro, Marcelo A},
  journal={Physical Review B},
  volume={68},
  number={13},
  pages={134401},
  year={2003},
  publisher={APS}
}

@article{bab2019evidence,
  title={Evidence of Kosterlitz-Thouless phase transitions in the Ising model with dipolar interactions},
  author={Bab, Marisa Alejandra and Saracco, Gustavo Pablo},
  journal={Physical Review E},
  volume={100},
  number={2},
  pages={022143},
  year={2019},
  publisher={American Physical Society}
}

@article{horowitz2015phase,
  title={Phase transitions and critical phenomena in the two-dimensional Ising model with dipole interactions: A short-time dynamics study},
  author={Horowitz, CM and Bab, Marisa Alejandra and Mazzini, Marcos and Puzzo, ML Rubio and Saracco, Gustavo Pablo},
  journal={Physical Review E},
  volume={92},
  number={4},
  pages={042127},
  year={2015},
  publisher={American Physical Society}
}

@article{PhysRevB.51.16033,
  title = {Striped phases in two-dimensional dipolar ferromagnets},
  author = {MacIsaac, A. B. and Whitehead, J. P. and Robinson, M. C. and De'Bell, K.},
  journal = {Phys. Rev. B},
  volume = {51},
  issue = {22},
  pages = {16033--16045},
  numpages = {0},
  year = {1995},
  month = {Jun},
  publisher = {American Physical Society},
  doi = {10.1103/PhysRevB.51.16033},
  url = {https://link.aps.org/doi/10.1103/PhysRevB.51.16033}
}

@article{arlett1996phase,
  title={Phase diagram for the striped phase in the two-dimensional dipolar Ising model},
  author={Arlett, J and Whitehead, JP and MacIsaac, AB and De’Bell, K},
  journal={Physical Review B},
  volume={54},
  number={5},
  pages={3394},
  year={1996},
  publisher={APS}
}

@article{8a265d0d-ebde-3389-9ee4-4c2c542a04ef,
 ISSN = {00368075, 10959203},
 URL = {http://www.jstor.org/stable/2886194},
 author = {Michael Seul and David Andelman},
 journal = {Science},
 number = {5197},
 pages = {476--483},
 publisher = {American Association for the Advancement of Science},
 title = {Domain Shapes and Patterns: The Phenomenology of Modulated Phases},
 urldate = {2026-08-26},
 volume = {267},
 year = {1995}
}

@article{PhysRevE.111.055406,
  title = {Active microrheology and dynamic phases for pattern-forming systems with competing interactions},
  author = {Reichhardt, C. and Reichhardt, C. J. O.},
  journal = {Phys. Rev. E},
  volume = {111},
  issue = {5},
  pages = {055406},
  numpages = {13},
  year = {2025},
  month = {May},
  publisher = {American Physical Society},
  doi = {10.1103/PhysRevE.111.055406},
  url = {https://link.aps.org/doi/10.1103/PhysRevE.111.055406}
}

@article{10.1063_1.328121,
    author = {D{\"o}tsch, H. and Tolksdorf, W. and Welz, F.},
    title = {Domain‐wall oscillation of bubble and stripe lattices in hexagonal ferrites},
    journal = {Journal of Applied Physics},
    volume = {51},
    number = {7},
    pages = {3816-3820},
    year = {1980},
    month = {07},
    issn = {0021-8979},
    doi = {10.1063/1.328121},
    url = {https://doi.org/10.1063/1.328121},
}

@article{kooy1960experimental,
  title={Experimental and theoretical study of the domain configuration in thin layers of {BaFe}$_{12}${O}$_{19}$},
  author={Kooy, C},
  journal={Philips Res. Repts},
  volume={15},
  number={7},
  year={1960}
}

@inproceedings{10.1117_12.961386,
author = {Peter Meyer and Jacques Pommier and Jacques Ferre},
title = {{Magnetooptic Observation Of Domains At Low Temperature In The Transparent Ferromagnet LiHoF[sub]4[/sub]}},
volume = {1126},
booktitle = {Electro-Optic and Magneto-Optic Materials and Applications},
editor = {Jean-Paul Castera},
organization = {International Society for Optics and Photonics},
publisher = {SPIE},
pages = {93 -- 98},
year = {1989},
doi = {10.1117/12.961386},
URL = {https://doi.org/10.1117/12.961386}
}

@article{battison1975ferromagnetism,
  title={Ferromagnetism in lithium holmium $\text{fluoride-LiHoF}_4$. $\text{II}$. Optical and spectroscopic measurements},
  author={Battison, JE and Kasten, A and Leask, MJM and Lowry, JB and Wanklyn, BM},
  journal={Journal of Physics C: Solid State Physics},
  volume={8},
  number={23},
  pages={4089--4095},
  year={1975}
}

@article{10.1063_1.357538,
    author = {Barnes, J. R. and O’Shea, S. J. and Welland, M. E. and Kim, J.‐Y. and Evetts, J. E. and Somekh, R. E.},
    title = {Magnetic force microscopy of Co‐Pd multilayers with perpendicular anisotropy},
    journal = {Journal of Applied Physics},
    volume = {76},
    number = {5},
    pages = {2974-2980},
    year = {1994},
    month = {09},
    issn = {0021-8979},
    doi = {10.1063/1.357538},
    url = {https://doi.org/10.1063/1.357538},
}

@article{LOUAIL1997387,
title = {Magnetic configurations in multilayers: Temperature dependent cone states and narrow stripe domains},
journal = {Journal of Magnetism and Magnetic Materials},
volume = {165},
number = {1},
pages = {387-390},
year = {1997},
note = {Symposium E: Magnetic Ultrathin Films, Multilayers and Surfaces},
issn = {0304-8853},
doi = {https://doi.org/10.1016/S0304-8853(96)00565-3},
url = {https://www.sciencedirect.com/science/article/pii/S0304885396005653},
author = {L. Louail and K. Ounadjela and M. Hehn and K. Khodjaoui and M. Gester and H. Danan and R.L. Stamps},
}

@article{tovaglieri2026superdomains,
  title={Superdomains and superdomain walls in ferroelectric thin films},
  author={Tovaglieri, Ludovica and Paruch, Patrycja and Triscone, Jean-Marc and Lichtensteiger, C{\'e}line},
  journal={Journal of Applied Physics},
  volume={139},
  number={15},
  year={2026},
  publisher={AIP Publishing}
}

@article{gabay1985phase,
  title={Phase transitions and size effects in the Ising dipolar magnet},
  author={Gabay, M and Garel, T},
  journal={Journal de physique},
  volume={46},
  number={1},
  pages={5--16},
  year={1985},
  publisher={Soci{\'e}t{\'e} fran{\c{c}}aise de physique}
}

@article{rizzi2010phase,
  title={Phase transitions and autocorrelation times in two-dimensional Ising model with dipole interactions},
  author={Rizzi, Leandro G and Alves, Nelson A},
  journal={Physica B: Condensed Matter},
  volume={405},
  number={6},
  pages={1571--1579},
  year={2010},
  publisher={Elsevier}
}

@article{principi2016stripe,
  title={Stripe glasses in ferromagnetic thin films},
  author={Principi, Alessandro and Katsnelson, Mikhail I},
  journal={Physical Review B},
  volume={93},
  number={5},
  pages={054410},
  year={2016},
  publisher={APS}
}

@article{booth1995domain,
  title={Domain structures in ultrathin magnetic films},
  author={Booth, I and MacIsaac, AB and Whitehead, JP and De'Bell, K},
  journal={Physical review letters},
  volume={75},
  number={5},
  pages={950},
  year={1995},
  publisher={APS}
}

@article{whitehead2008canted,
  title={Canted stripe phase near the spin reorientation transition in ultrathin magnetic films},
  author={Whitehead, JP and MacIsaac, AB and De’Bell, K},
  journal={Physical Review B—Condensed Matter and Materials Physics},
  volume={77},
  number={17},
  pages={174415},
  year={2008},
  publisher={APS}
}

@article{simmons2024ferromagnetic,
  title={Ferromagnetic ordering in mazelike stripe liquid of a dipolar six-state clock model},
  author={Simmons, Quentin and Lin, Shi-Zeng and Chern, Gia-Wei},
  journal={arXiv preprint arXiv:2412.09550},
  year={2024}
}

@article{ewald1921ewald,
  title={Ewald summation},
  author={Ewald, Paul Peter},
  journal={Ann. Phys},
  volume={369},
  number={253},
  pages={1--2},
  year={1921}
}

@article{lekner1989summation,
  title={Summation of dipolar fields in simulated liquid-vapour interfaces},
  author={Lekner, John},
  journal={Physica A: Statistical Mechanics and its Applications},
  volume={157},
  number={2},
  pages={826--838},
  year={1989},
  publisher={Elsevier}
}

@article{sampaio1996magnetic,
  title={Magnetic relaxation and formation of magnetic domains in ultrathin films with perpendicular anisotropy},
  author={Sampaio, LC and De Albuquerque, MP and De Menezes, FS},
  journal={Physical Review B},
  volume={54},
  number={9},
  pages={6465},
  year={1996},
  publisher={APS}
}

@article{cannas2008interplay,
  title={Interplay between coarsening and nucleation in an Ising model with dipolar interactions},
  author={Cannas, Sergio A and Michelon, Mateus F and Stariolo, Daniel A and Tamarit, Francisco A},
  journal={Physical Review E—Statistical, Nonlinear, and Soft Matter Physics},
  volume={78},
  number={5},
  pages={051602},
  year={2008},
  publisher={APS}
}

@article{hosiawa2009late,
  title={Late stage, non-equilibrium dynamics in the dipolar Ising model},
  author={Hosiawa, Tom and MacIsaac, AB},
  journal={Journal of magnetism and magnetic materials},
  volume={321},
  number={12},
  pages={1878--1884},
  year={2009},
  publisher={Elsevier}
}

@article{bromley2003memory,
  title={Memory effects and slow dynamics in ultra thin magnetic films},
  author={Bromley, SP and Whitehead, JP and De'Bell, K and MacIsaac, AB},
  journal={Journal of magnetism and magnetic materials},
  volume={264},
  number={1},
  pages={14--29},
  year={2003},
  publisher={Elsevier}
}

@article{de2020emergent,
  title={Emergent magnetic structures and dynamics in thin films: a review of some recent results},
  author={De’Bell, K},
  journal={Canadian Journal of Physics},
  volume={98},
  number={9},
  pages={825--833},
  year={2020},
  publisher={NRC Research Press 1840 Woodward Drive, Suite 1, Ottawa, ON K2C 0P7}
}

@article{thompson2008imaging,
  title={Imaging and alignment of nanoscale 180 stripe domains in ferroelectric thin films},
  author={Thompson, Carol and Fong, DD and Wang, RV and Jiang, F and Streiffer, SK and Latifi, K and Eastman, JA and Fuoss, PH and Stephenson, GB},
  journal={Applied Physics Letters},
  volume={93},
  number={18},
  year={2008},
  publisher={AIP Publishing}
}

@article{tripathi2019coarsening,
  title={Coarsening dynamics in the Swift-Hohenberg equation with an external field},
  author={Tripathi, Ashwani K and Kumar, Deepak and Puri, Sanjay},
  journal={Physical Review E},
  volume={99},
  number={2},
  pages={022136},
  year={2019},
  publisher={APS}
}

@article{gupta2025anisotropic,
  title={Anisotropic spin stripe domains in bilayer $\text{La}_3\text{Ni}_2\text{O}_7$},
  author={Gupta, Naman K and Gong, Rantong and Wu, Yi and Kang, Mingu and Parzyck, Christopher T and Gregory, Benjamin Z and Costa, Noah and Sutarto, Ronny and Sarker, Suchismita and Singer, Andrej and others},
  journal={Nature Communications},
  volume={16},
  number={1},
  pages={6560},
  year={2025},
  publisher={Nature Publishing Group UK London}
}

@article{sun2024reversible,
  title={Reversible Mechanical Switching of Ferroelastic Stripe Domains in Multiferroic Thin Films},
  author={Sun, Fei and Wu, Mengjun and Ren, Jianhua and Wang, Xintong and Yang, Hui and Zhang, Xiaoyue and Chen, Weijin and Zheng, Yue},
  journal={ACS Applied Materials \& Interfaces},
  volume={16},
  number={25},
  pages={32425--32433},
  year={2024},
  publisher={American Chemical Society}
}

@article{saxena2025strain,
  title={Strain-driven Domain Wall Network with Chiral Junctions in an Antiferromagnet},
  author={Saxena, Vishesh and Gutzeit, Mara and Rodr{\'\i}guez-Sota, Arturo and Haldar, Soumyajyoti and Zahner, Felix and Wiesendanger, Roland and Kubetzka, Andr{\'e} and Heinze, Stefan and von Bergmann, Kirsten},
  journal={Nature Communications},
  volume={16},
  number={1},
  pages={10808},
  year={2025},
  publisher={Nature Publishing Group UK London}
}

@article{moon2019measuring,
  title={Measuring the Magnetization from the Image of the Stripe Magnetic Domain},
  author={Moon, Kyoung-Woong and Yoon, Jungbum and Choi, Jun Woo and Kim, Changsoo and Kim, Dong-Ok and Kim, Dongseuk and Chun, Byong Sun and Min, Byoung-Chul and Hwang, Chanyong},
  journal={Physical Review Applied},
  volume={12},
  number={3},
  pages={034030},
  year={2019},
  publisher={APS}
}

@article{shi2023domain,
  title={Domain-dependent Strain and Stacking in Two-Dimensional van der Waals Ferroelectrics},
  author={Shi, Chuqiao and Mao, Nannan and Zhang, Kena and Zhang, Tianyi and Chiu, Ming-Hui and Ashen, Kenna and Wang, Bo and Tang, Xiuyu and Guo, Galio and Lei, Shiming and others},
  journal={Nature Communications},
  volume={14},
  number={1},
  pages={7168},
  year={2023},
  publisher={Nature Publishing Group UK London}
}

@article{singh2018nucleation,
  title={Nucleation of Stripe Domains in Thin Ferromagnetic Films},
  author={Singh, S and Gao, H and Hartmann, U},
  journal={Physical Review B},
  volume={98},
  number={6},
  pages={060414},
  year={2018},
  publisher={APS}
}

@article{huh2024universality,
  title={Universality Class of a Spinor Bose--Einstein Condensate Far from Equilibrium},
  author={Huh, SeungJung and Mukherjee, Koushik and Kwon, Kiryang and Seo, Jihoon and Hur, Junhyeok and Mistakidis, Simeon I and Sadeghpour, Hossein R and Choi, Jae-yoon},
  journal={Nature Physics},
  volume={20},
  number={3},
  pages={402--408},
  year={2024},
  publisher={Nature Publishing Group UK London}
}

@article{fumeron2023introduction,
  title={Introduction to Topological Defects: From Liquid Crystals to Particle Physics},
  author={Fumeron, S{\'e}bastien and Berche, Bertrand},
  journal={The European Physical Journal Special Topics},
  volume={232},
  number={11},
  pages={1813--1833},
  year={2023},
  publisher={Springer}
}

@article{angheluta2026full,
  title={Full-Integer Topological Defects in Polar Active Matter: From Collective Migration to Tissue Patterning},
  author={Angheluta, Luiza and L{\aa}ng, Anna and L{\aa}ng, Emma and B{\o}e, Stig Ove},
  journal={Annual Review of Condensed Matter Physics},
  volume={17},
  year={2026},
  publisher={Annual Reviews}
}

@article{balch2023spatially,
  title={Spatially extended dislocations produced by the dispersive Swift-Hohenberg equation},
  author={Balch, Brenden and Shipman, Patrick D and Bradley, R Mark},
  journal={Physical Review E},
  volume={107},
  number={4},
  pages={044214},
  year={2023},
  publisher={APS}
}

@article{singh2021late,
  title={Late Stage Domain Coarsening Dynamics of Lamellar Block Copolymers},
  author={Singh, Maninderjeet and Wu, Wenjie and Nuka, Vinay and Strzalka, Joseph and Douglas, Jack F and Karim, Alamgir},
  journal={ACS Macro Letters},
  volume={10},
  number={6},
  pages={727--731},
  year={2021},
  publisher={ACS Publications}
}

@article{tripathi2015coarsening,
  title={Coarsening of Stripe Patterns: Variations with Quench Depth and Scaling},
  author={Tripathi, Ashwani K and Kumar, Deepak},
  journal={Physical Review E},
  volume={91},
  number={2},
  pages={022923},
  year={2015},
  publisher={APS}
}

@article{liang2026phase,
  title={Phase-Field Simulations in Ferroic Nanodomains: Defect-Field Driven Microstructure--Functionality Manipulation},
  author={Liang, Chuanxin and He, Liqiang and Zhang, Le and Wang, Dong},
  journal={Advanced Materials},
  volume={38},
  number={29},
  pages={e15831},
  year={2026},
  publisher={Wiley Online Library}
}

@article{indergand2023domain,
  title={Domain Pattern Formation in Tetragonal Ferroelectric Ceramics},
  author={Indergand, Roman and Bruant, Xavier and Kochmann, Dennis M},
  journal={Journal of the Mechanics and Physics of Solids},
  volume={181},
  pages={105426},
  year={2023},
  publisher={Elsevier}
}

@article{pinna2025mechanisms,
  title={Mechanisms of Alignment of Lamellar-Forming Block Copolymer under Shear Flow},
  author={Pinna, Marco and Diaz, Javier and Denison, Christopher and Zvelindovsky, Andrei and Pagonabarraga, Ignacio},
  journal={Soft Matter},
  volume={21},
  number={3},
  pages={476--487},
  year={2025},
  publisher={The Royal Society of Chemistry}
}

@article{patel2022rapid,
  title={Rapid, Interface-Driven Domain Orientation in Bottlebrush Diblock Copolymer Films during Thermal Annealing},
  author={Patel, Bijal B and Walsh, Dylan J and Patel, Kush and Kim, Do Hoon and Kwok, Justin J and Guironnet, Damien and Diao, Ying},
  journal={Soft Matter},
  volume={18},
  number={8},
  pages={1666--1677},
  year={2022},
  publisher={The Royal Society of Chemistry}
}

@article{leniart2022pathway,
  title={Pathway-Dependent Grain Coarsening of Block Copolymer Patterns under Controlled Solvent Evaporation},
  author={Leniart, Arkadiusz A and Pula, Przemyslaw and Style, Robert W and Majewski, Pawel W},
  journal={ACS Macro Letters},
  volume={11},
  number={1},
  pages={121--126},
  year={2022},
  publisher={ACS Publications}
}

@article{blagojevic2023multiscale,
  title={Multiscale Modelling of Grain-Boundary Motion in Cylinder-Forming Block Copolymers},
  author={Blagojevic, Niklas and M{\"u}ller, Marcus},
  journal={ACS Polymers Au},
  volume={3},
  number={1},
  pages={96--117},
  year={2023},
  publisher={ACS Publications}
}

@article{mondal2024ordering,
  title={Ordering Kinetics and Steady States of the XY Model with Ferromagnetic and Nematic Interaction},
  author={Mondal, Partha Sarathi and Mishra, Pawan Kumar and Mishra, Shradha},
  journal={Journal of Physics: Condensed Matter},
  volume={36},
  number={28},
  pages={285101},
  year={2024},
  publisher={IOP Publishing}
}




\end{document}